\pdfoutput=1
\documentclass[twocolumn,showpacs,amssymb,floatfix,deluxe,prd]{revtex4}

\usepackage{graphicx}% Include figure files
\usepackage{dcolumn}% Align table columns on decimal point
\usepackage{bm}% bold math
\usepackage{epsfig}
\usepackage[english]{babel}
\usepackage{amsmath}
\usepackage{amssymb}

\usepackage{verbatim}
\usepackage{alltt}

\def\({\left(}
\def\){\right)}

\begin{document}
CERN-PH-TH/2010-120
\preprint{CERN-PH-TH/2010-120}
\title{
%% old 
%% new
Measuring the W-Boson mass at a hadron collider:\\
a study of phase-space singularity methods} 
% Force line breaks with \\

\author{A. De R\'ujula${}^{a,b,c}$}
\affiliation{  \vspace{3mm}
${}^a$Instituto de F\'isica Te\'orica (UAM/CSIC), Univ. Aut\'onoma de Madrid, Madrid, and 
CIEMAT, Madrid, Spain,\\
${}^b$Physics Dept., Boston University, Boston, MA 02215,\\
${}^c$Physics Department, CERN, CH 1211 Geneva 23, Switzerland}

\author{A. Galindo${}^{d,e}$}
\affiliation{${}^d$Departamento de F\'isica,
Universidad Complutense, Madrid, Spain, ${}^e$CIEMAT, Madrid, Spain}%

\date{\today}% It is always \today, today,

\begin{abstract}

The traditional method to measure the $W$ Boson mass at a hadron collider 
(more precisely, its ratio to the $Z$ boson mass) utilizes the distributions of three variables
in events where the $W$ decays into an electron or a muon: the 
charged lepton transverse momentum, the missing transverse energy and
the transverse mass of the lepton pair. We study the  putative advantages of
the additional measurement of a fourth variable: an improved phase space singularity mass.
This variable is statistically optimal, and simultaneously exploits the longitudinal- and
transverse-momentum distributions of the charged lepton. 
Though the process we discuss is one of the simplest realistic ones involving just one
unobservable particle, it is fairly nontrivial and constitutes a good ``training"
example for the scrutiny of phenomena involving invisible objects. Our graphical analysis 
of the phase space is akin to that of a Dalitz plot, extended to such processes.
\end{abstract}

\maketitle

\section{Prolegomena}

Neutrinos --and perhaps novel weakly-interacting particles-- escape unobserved
from the collisions in which they are produced. In the corresponding ``missing energy"
events, the reconstruction of the masses
of the parent particles and the specification of the underlying process are
challenging because there are typically fewer kinematical constraints
than unknowns. At a hadron collider this situation is rendered even thornier,
since particles produced at small angles also escape undetected. This prohibits
the determination of the longitudinal momentum
of the center of mass system of the colliding partons. 

The above limitations confer a higher standing to observables exclusively dependent on 
transverse momenta \cite{mT2}, or otherwise invariant under longitudinal boosts \cite{CR}.
 In principle, transverse observables are 
 insensitive to the significant uncertainties associated
with the (longitudinal) parton distribution functions (pdfs).
In practice the uncertainties are to some extent reintroduced via
the angular coverage limitations
of  an actual experiment, which are not invariant under longitudinal boosts.

The quintessential transverse observable is the transverse mass,
of $W$-discovery fame. 
In an event at a hadron collider, consider the production of a single $W$,
followed by its decay $W\to \ell \nu$, with $\ell$ an electron, a muon, or one of
their antiparticles. Denote by $x\equiv (x_0,\vec x_{_T},x_3)$ 
and $l\equiv (l_0,\vec l_{_T},l_3)$ the neutrino and charged lepton
fourmomenta, respectively. Here $\vec l_{_T}\equiv (l_1,l_2)$ and 
$\vec x_{_T}\equiv (x_1,x_2)$ are the momenta of the leptons
 in the plane transverse to the beam direction(s), and
$\vec p_{_T}\equiv (p_1,p_2)$ the analogous quantity for the observed  final state hadrons.
The traditional ``transverse mass", a function of $\vec l_T$ and $\vec p_T$, 
whose distribution is used to infer the $W$ boson mass, is \cite{mT2}
\begin{eqnarray}
M_{_T}^2  &=& 2\,l_{_T}\,x_{_T}\,[1-\cos\Delta\Phi(\vec x_{_T},\vec l_{_T})]\nonumber\\
\vec x_{_T}&\backepsilon& \vec x_{_T}+ \vec l_{_T} + \vec p_{_T} = 0,
\label{mT2}
\end{eqnarray}
where $\Delta\Phi(\vec x_{_T},\vec l_{_T})$ is the angle between the transverse
lepton directions.
The most precise determination of the mass of the $W$ 
by a single experiment is the one by D$\O$ \cite{D0}.
In spite of the relatively unfavorable environment of a hadron collider,
its large statistics results in a value with an overall error smaller than 
that of the LEP experiments.
The D$\O$ result is based on the decays $W\to e\,\nu$, and
the measurement of three highly correlated transverse observables:
the traditional ``transverse mass" function \cite{mT2}, the lepton's transverse
energy and the total missing transverse energy. The result:
\begin{equation}
M_W=80.401\pm 0.043\; \rm GeV,
\label{MW}
\end{equation}
stems from an actual measurement of $M_W/M_Z$. 
But $M_Z$ was determined with exquisite precision
at LEP. The PDG quotes $M_Z=91.1876 \pm 0.0021$ GeV \cite{pdg}. 

The procedure to extract $M_W$ from the
distributions in transverse mass, lepton momentum and
 total missing energy is as follows.
A finely spaced set of input $W$ boson masses, $M$, is used to generate a set
of ``templates": the ``Monte Carlo" (MC) expectations for the observed distributions,
with all their experimental cuts, estimated uncertainties, calorimeter responses, etc.
The $\chi^2({ M})$ values for the comparison of data and expectations
are fit to a quadratic form, from whose minimum and width
 $M_W$ and its estimated error are inferred.
Naturally, all the procedure is tested and calibrated by the observed $Z$-production
and leptonic decay (into $e^+ e^-$, in the D$\O$ case).

In order of decreasing incidence on the error in  Eq.~({\ref{MW}), the limitations
are the electron's energy calibration, the uncertainties on 
the pdfs, and the statistics. For this particular measurement, the backgrounds are well
understood and quite negligible. 

Given the large statistics already gathered at the Tevatron collider, and
with the advent of the LHC as a high statistics precision physics tool,
the main limitation of a hadron collider determination of the $W$ mass from its decays into
electrons and muons is likely to be the pdf uncertainty. 
At the LHC this problem is in particular exacerbated \cite{Dydak}
by the fact that it is a $pp$, not a $\bar p p$ collider, and
the quark pdfs in a proton --or the
 identical antiquark pdfs in an antiproton-- 
are much better known than the antiquark pdfs in a proton.

\section{Introduction}
\label{sec:I}

A ginormous amount of attention has been paid to hypothetical processes
involving neutral, long-lived, weakly-interacting final state particles that 
can only be indirectly detected. A prototypical example is the pair production
of squarks followed by their decays into quark plus neutralino. Such processes
generally involve two or more particles of unknown masses. 

The first aim in the missing particle searches for physics beyond the
Standard Model is the establishment or the exclusion of a signal, 
both tantamount to an efficient suppression of backgrounds. 
Some novel longitudinal boost invariant variables are a very good choice
in this endeavor \cite{CR}, as demonstrated by the data analysis in
\cite{CMS}.

A longer range aim is the measurement of unknown masses, when there
are more than one and a candidate process is selected. In this connection,
a very general algebraic singularity method has been advocated \cite{Kim}, 
involving the use of a ``singularity variable" (SV), allegedly more powerful than
that of a singularity ``condition" (SC), such as the one leading, as we shall see, 
to the $M_{_T}^2$ result of Eq.~(\ref{mT2}). 

It is too late to discover the $W$, though not to attempt to measure its mass even
better, a relevant task in checking the consistency of the Standard Model
and constraining the mass of its hypothetical scalar. With this ab-initio motivation,
we have exhaustively studied the phase space for $W$ production and leptonic
decay, a simple undertaking analogous to the analysis of a Dalitz plot, but with
 incomplete kinematical information (\S \ref{sec:SC}).

We have also studied the singularities 
of this phase space, and their use in constraining the $W$ mass
(\S \ref{sec:SC} and \ref{sec:KSV}) . We identify
the criterion for the theoretically optimal SV and derive its explicit form
(\S \ref{sec:QOV}, \ref{sec:R} and \ref{sec:GC}). En passant, we
find that other nonoptimal SVs, such as the one proposed in  \cite{Kim},
are ``dangerous", in that their distributions display fake singularities
(\S \ref{sec:IS}).
 
The singularity variables we study involve the measured longitudinal
momentum of the charged lepton, $l_3$.
This longitudinal information 
 is obviously additive to the transverse information exploited in observables such as $M_T^2$,
 but is highly correlated with it (\S \ref{sec:C}).
The $l_3$ distribution directly reflects the pdfs of merging quarks and antiquarks
of different flavor. Recent progress in QCD fits and in calculations well
beyond the leading order allows one to hope that --eventually-- the dominant limitations
concerning the problem at hand will not be the theoretical pdf uncertainties,
but the limited calorimetric resolutions. 

Given a trustable set of pdfs, one can  simulate
the observable distribution of events $dN/(dl_3\,d^2 l_{_T}\,d^2 p_{_T})$ for a set
of input trial masses and contrast it with observation. This comparison involves
the five relevant variables and their correlations; it has no statistically superior
competitor. Why then study any alternatives? Besides the pleasure of understanding
with use of one's own neural network, there is the motivation of paving
the way of searches for other processes involving unobservable particles, for
which it is a-priori prohibitive to simulate all possibilities.

In this note we report on a thorough theoretical study of the extraction
of phase space information from single-$W$ signal events, 
but we use the standard model of 
$W$ production and decay only to leading order. We entirely ignore the backgrounds,
which are well known to be very modest for this particular process.
A reason for these choices is that only the experimentalists
themselves can fully model the detector's effects and backgrounds, and that this
modeling is  independent from the theoretical issues on which
we focus.

\section{Linguistic quandaries}

Based on equations such as $M^2=(l+x)^2$, we shall be drawn to give
a plethora of meanings to what is, for starters, simply a letter: $``M"$.
It ends up being everything else. The resemblance to $M$-theory is 
coincidental.

Naturally, $M$ may stand for the physical or measured $M_W$, as
well as for its Lorentzian distribution, when the width is not neglected.
But it may also, as in the case of the transverse mass, $M_T$,
be a non-Lorentzian function of other observables. 

In analyzing data, one compares them with MC generated distributions that depend
on an ensemble of input ``trial masses", for which we reserve the label ${ M}$.
A different type of trial masses, which we call $\cal M$, appears in ``singularity {\it variables"},
which are {\it functions} of observable momenta
{\it and} of $\cal M$. Not to make this complex linguistic heritage hereditary,
we label the singularity variables $``\Sigma"$ (and not once more ``$M$", as in the
$M_T^2$ function)
thereby not introducing new meanings to the symbol $M$ or the word ``mass".

\section{Single-$W$ phase space}
\label{sec:SC}

The full information relevant to the reconstruction of the $W$
mass is embedded in the kinematical equations:
\begin{eqnarray}
&&E_1\Rrightarrow x^2=0\label{singleW1}\\
&&E_2\Rrightarrow 2\;l\cdot x =M^2\label{singleW2}\\
&&E_3\Rrightarrow l_1+x_1+p_1=0\label{singleW3}\\
&&E_4\Rrightarrow l_2+x_2+p_2=0
\label{singleW4}
\end{eqnarray}
where we have made the approximation $l^2=0$ for the charged lepton.
The equations are incomplete in that  the $\nu$ longitudinal 
momentum, $x_3$, is unconstrained, precluding a direct determination of 
the $W$ boson mass from a ``mass peak". Is there a systematic way 
to extract the kinematically most stringent information on $M_W$?

To answer this question it is useful to study first the phase space described by
Eqs.(\ref{singleW1}-\ref{singleW4}) in a simplified case.
If the energy and transverse momentum of the observed
 hadrons could be measured with precision, it would be possible
to boost every event to the $\vec p_{_T}=0$ frame. To (temporarily) simplify the
algebra, let us just adopt this constraint. Solve the linear equations $E_2,E_3,E_4$
to express $x_0,x_1,x_2$ as functions of $x_3$. Substitute the result in $E_1$ to
obtain the phase space
%, a surface in $(l_1,l_2,l_3,x_3)$ space:
\begin{eqnarray}
&&\!\!\!\!\Phi(l_T,l_3,x_3,M)\equiv
\nonumber\\
&&(  M^2 + 2\, l_3 \,x_3-2\, l_{_T}^2)^2 - 4 \,l_0^2 \,(l_{_T}^2 + x_3^2)=0
\label{PhaseSp0}\\
&&\!\!\!\! l_0\equiv +\,\sqrt{l_{_{T}}^2+l_3^2}\label{PhaseSp1}\\
&&\!\!\!\!  l_{_T}^2\equiv {l_1^2+l_2^2}
 \label{PhaseSp2}
\end{eqnarray}
It will be useful to consider the two solutions to Eq.(\ref{PhaseSp0}) in 
$x_3=x_3(l_T,l_3,M)$:
\begin{equation}
x_3^\pm={1\over 2\,l_T^2}\left[l_3(M^2-2\,l_T^2)\pm M\,l_0\sqrt{M^2-4\,l_T^2}\right]
\label{x3solution}
\end{equation}

With no loss of generality, and to be able to plot the phase space, do three
more things. Take $l_3$ to be positive if directed
along the direction of a given (fixed) proton beam. Define 
the $l_T$ of Eq.~(\ref{PhaseSp2}) to be positive if directed above the beams,
negative otherwise.
The function $\Phi(l_T,l_3,x_3)=0$, from divers points of view,
is plotted in Fig.~\ref{fig:phasespace}.
Along the (blue) straight lines the planes tangent to the phase space contain one
``visible" direction, $l_3$, and the ``invisible" direction $x_3$. The projection
of phase space into the visible directions $(l_T,l_3)$ is bounded 
by the lines $l_T=\pm M/2$.

The boundaries of the phase space projected along an
invisible direction onto the space of the visible ones,  $l_T^2=M^2/4$, are
an example of {\it singularity condition(s)}. At their location there is a single invisible
coordinate $x_3$ for fixed values $(l_T,l_3)$ of the visible ones, as opposed to the 
two of the general case in Eq.~(\ref{x3solution}), and the projected phase space
density is not smooth  \cite{Kim}.

\begin{figure}[htbp]
\begin{center}
\includegraphics[width=0.4\textwidth]{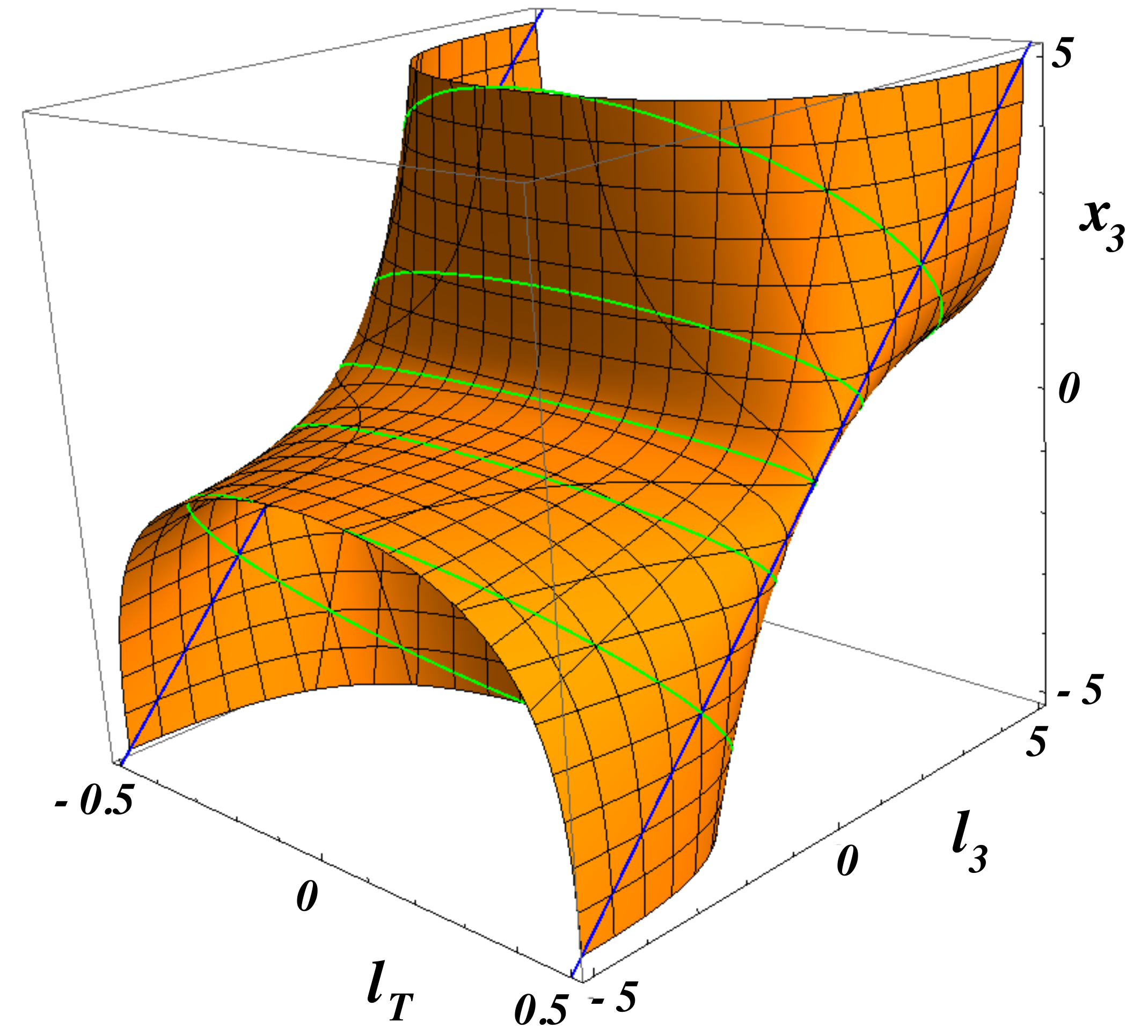}\\
\includegraphics[width=0.4\textwidth]{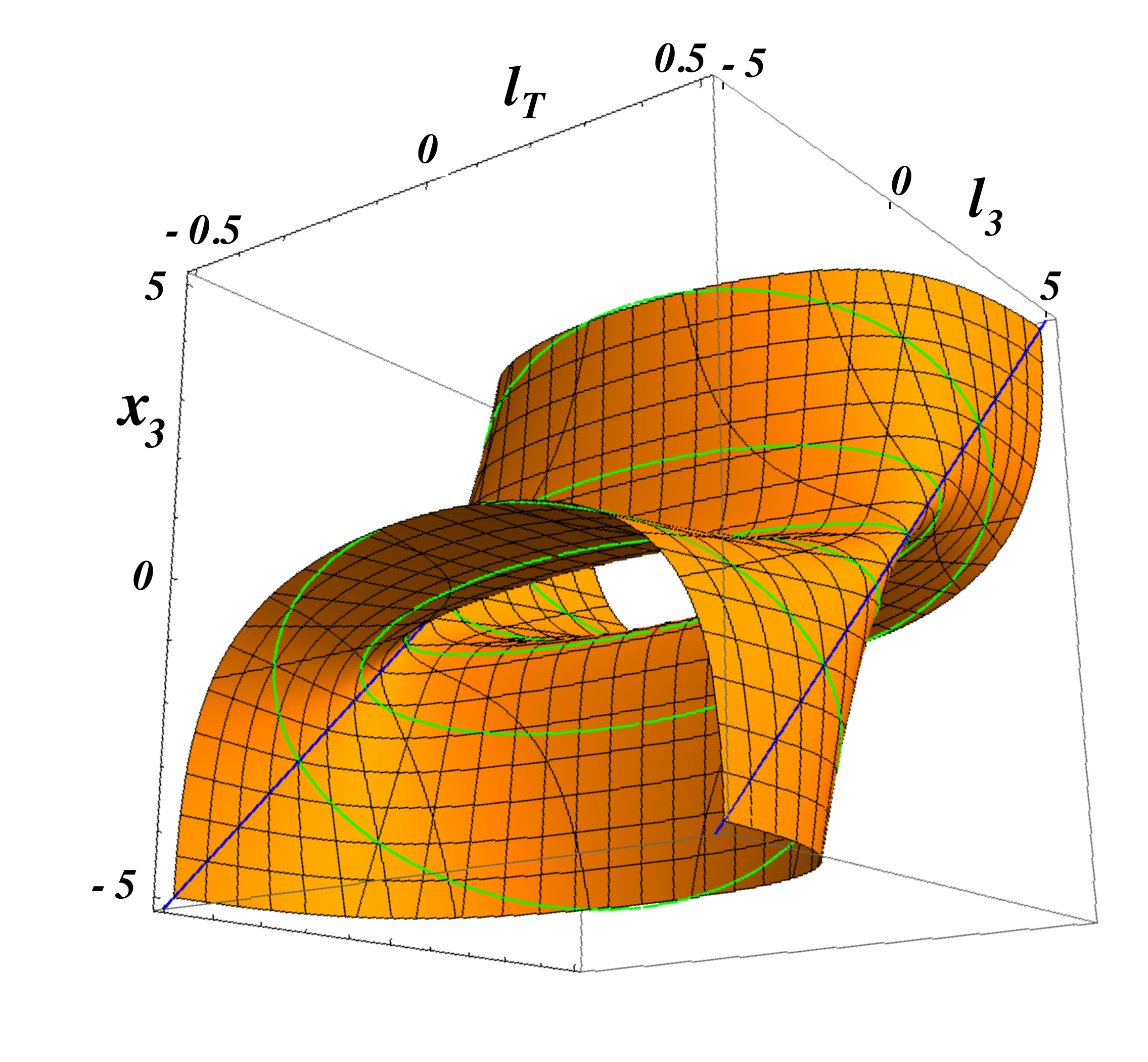}\\
\includegraphics[width=0.36\textwidth]{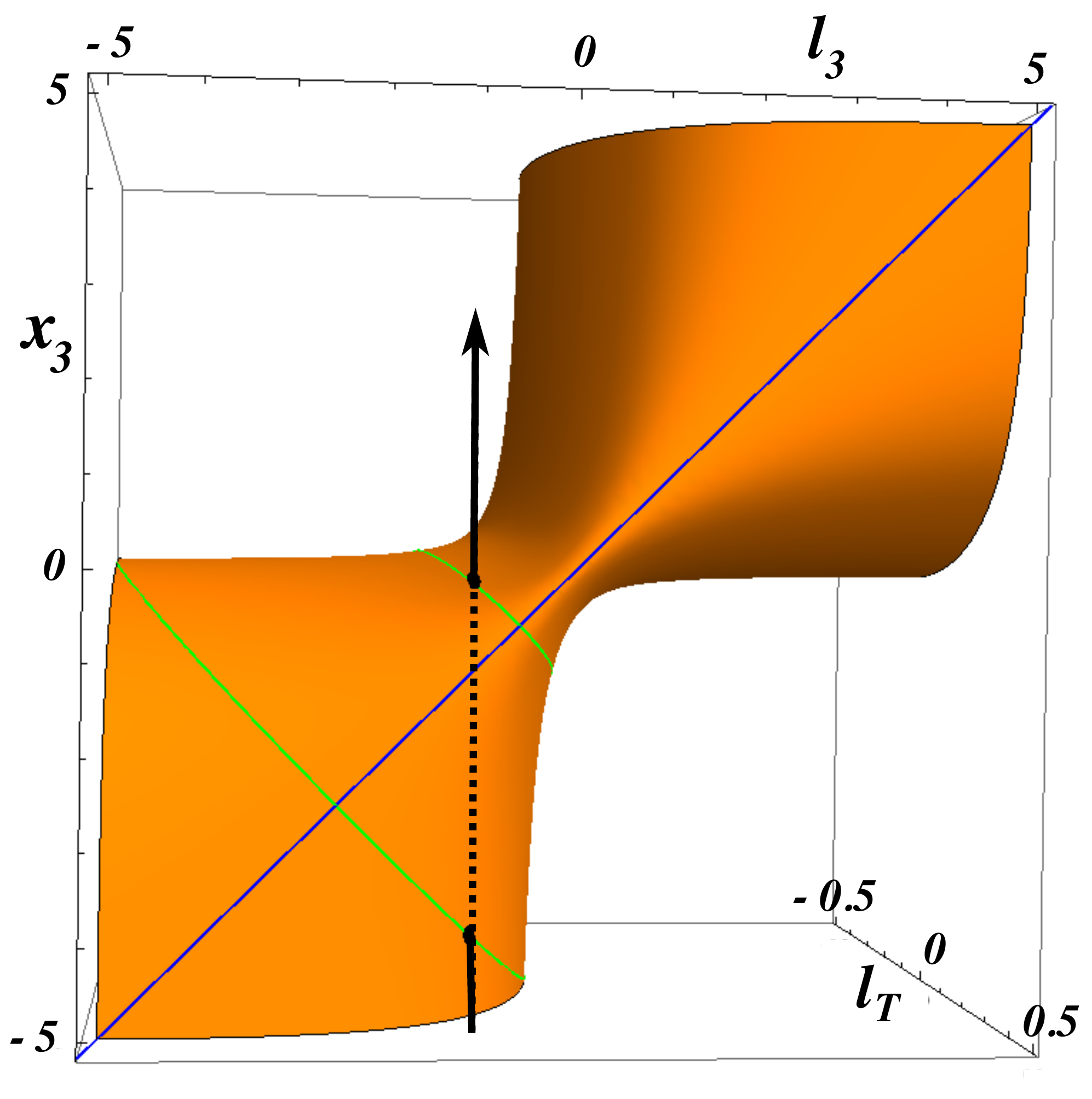}\\
\caption{
Three views of the phase space function $\Phi$ of Eq.~(\ref{PhaseSp0}),
with the momenta ($l_T$, $l_3$ and $x_3$) in units of $M$.
The black lines cut the surface at fixed $l_T$ or $l_3$ and the green ellipses
at fixed $W_3=l_3+x_3$, the longitudinal momentum of the $W$.
The (blue) lines at $l_T=\pm 1/2$, $x_3=l_3$ are singular.
A point in the $(l_T,l_3)$ plane corresponds to two values
of $x_3=x_3^\pm (l_T,l_3)$. 
  \label{fig:phasespace}}
\end{center}
\end{figure}

In practice two cuts have to be applied to the momentum of the
observed lepton. We adopt $|l_3|<5\,|l_T|$ (resulting from 
a pseudo-rapidity limitation $|\bar \eta |< 2.3$) and 
a rather demandingly low $|l_T|>10$ GeV. 
These cuts result in the unobservability of
a large fraction of phase space: the (red) domain shown without
a mesh in Fig.~\ref{fig:phasespace2}. 
The maximum $|x_3|={\cal{O}}(50)\,M_W$ happens to be close to
the absolute kinematical limit, approximately $|x_3|< E_p$, at the current
LHC energy, $E_p=3.5$ TeV. This was probably not the
main reason to choose this machine energy.

\begin{figure}[htbp]
\begin{center}
\includegraphics[width=0.45\textwidth]{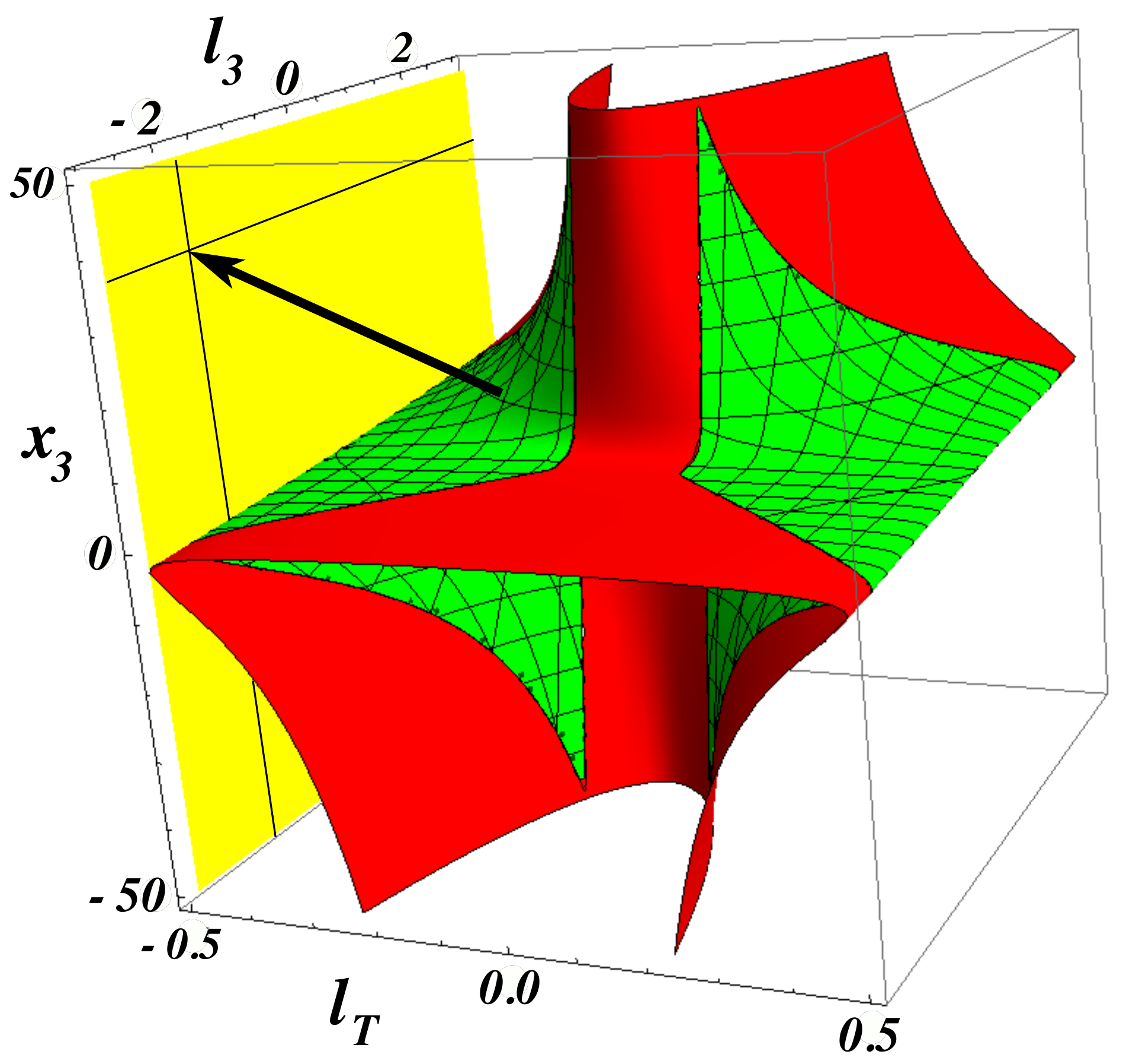}\\
\caption{
 The same as Fig.~\ref{fig:phasespace}, 
but in a different, more extensive, domain of $(l_T,l_3,x_3)$.
The finite dashed (green) domain is what survives the typical
experimental cuts on $l_T$ and $\bar \eta$. A (yellow) plane tangent
to the phase space surface $\Phi=0$ along the singularity line
at $l_T/M=-1/2$ is shown at the left; it contains the invisible direction $x_3$. 
The arrow is orthogonal to 
the phase space $\Phi=0$ at a point in it, and extends from this point to the tangent plane.
\label{fig:phasespace2}}
\end{center}
\end{figure}

In simple cases such as the one at hand the singularity condition can be 
directly obtained.  The $l_T$ boundary is the projection of the phase space
points at which the tangent plane is vertical and contains the invisible
direction $x_3$. At these points $\partial \Phi(l_T,l_3,x_3)/\partial x_3=0$.
Eliminating $M$ from this expression and Eq.~(\ref{PhaseSp0}) one
obtains $x_3= l_3$. At these boundaries $M^2=4\,l_T^2$.

\subsection{The formal singularity condition}

The procedure of the last paragraph requires some guesswork, but
can be rendered entirely general and systematic.
At a singularity one or more of the invisible directions are contained 
in the tangent plane to the full phase space.
The general condition for this to happen
is that, in the space $\{x\}$ of invisible directions, 
the row vectors of the Jacobian matrix
$D_{i j}\equiv\partial E_i/\partial x_j$ (with the row index $i$ running along the number of
equations and the column index
$j$ over the number of invisible coordinates) be linearly dependent,
so that the derivative relative to an $x$-direction normal to these vectors be zero. In other
words, at a singularity, the rank of $D_{i j}$ must be smaller than its rank at 
nonsingular points \cite{Kim}.

For the general single-$W$ case we are discussing
\begin{equation}
D\!=\!{\partial (E_1,E_2,E_3,E_4)\over \partial (x_0,x_1,x_2,x_3)}
\!=\!2 \!\left(
\begin{array}{cccc}
   {x_0} & -  {x_1} & -  {x_2} & -  {x_3} \\
   {l_0} & -  {l_1} & -  {l_2} & -  {l_3} \\
 0 & 1 & 0 & 0 \\
 0 & 0 & 1 & 0
\end{array}
\right)
\label{Dij}
\end{equation}
and the reduced rank condition is
\begin{equation}
E_C\Rrightarrow {\rm Det}\,D\propto l_0\,x_3-l_3\,x_0=0
\label{Det}
\end{equation}
The same condition is obtained in the $\vec p_T=0$ example.
Combining it with Eq.~(\ref{PhaseSp0}) results in $x_3=l_3$,
the phase space boundaries shown as straight (blue) lines in
Fig.~(\ref{fig:phasespace}).

\subsection{The $M_{_T}$ function}

The general case with nonvanishing $\vec p_{_T}$ is treated with equal ease.
Eliminate the four variables $x$ to solve the five equations 
(\ref{singleW1}-\ref{singleW4},\ref{Det}) in $M$. The result is $\Sigma_T=0$, with:
\begin{eqnarray}
\!\!\!\!\!\!\!\!\!\!&&\Sigma_T(M,\vec l_{_T},\vec p_{_T})\equiv \nonumber\\
\!\!\!\!\!\!\!\!\!\!&&M^4-4\,M^2\,(\vec l_{_{T}}\cdot \vec p_{_{T}}+l_{_{T}}^2)
+4\,\left[(\vec l_{_{T}} \cdot \vec p_{_{_T}})^2-l_{_{T}}^2\,p_{_{T}}^2\right]
\label{calMeq2}
\end{eqnarray}
Of the four $M$-roots of
$\Sigma_T=0$, one is not unphysical
\begin{equation}
{M_T}(\vec l_{_T},\vec p_{_T})
\!=\!+\sqrt{2\,\left[ |l_{_T}|\,|p+l|_{_T}+\vec l_{_T}\cdot (\vec l_{_T}+\vec p_{_T})\right]},
\label{mT2bis}
\end{equation}
which reduces to ${M_T} = 2\, |l_{_T}|$ for $\vec p_{_T}=0$.
The function ${M_T}^2$ of Eq.~(\ref{mT2bis}) is the consuetudinary $M_{_T}^2$ of
Eq.~(\ref{mT2}).

\section{Kim's singularity variable}
\label{sec:KSV}

Discussing the general case with an arbitrary number of invisible final state
particles, 
Kim has argued \cite{Kim} that the use of a ``singularity variable" (SV) is more powerful than
that of a singularity ``condition" (SC), such as the one leading to the $M_{_T}^2$
result of Eq.~(\ref{mT2bis}).

Kim requires a SV to have four properties \cite{Kim}:
\\
(i) To vanish at the singularity. 
\\
(ii)  To be perpendicular --at the singularity-- to the phase space surface in the observable
directions.
\\
(iii) To be ``normalized such that every event can give the same significance".\\
(iv) To be computed to first nontrivial order (the second fundamental
form) in the distance between a phase space 
point and the nearest singularity. 

Our interpretation of these formal looking choices is the following.
Condition (i) is the only scale invariant stipulation. 
At the singularity, condition (ii) entails a maximal sensitivity to
the unknown masses. Condition (iii) ensures that two events with the
same distance to the singularity be treated on equal footing.
The requirement (iv) is one way to make the procedure general.

To fathom all this it is useful to jump momentarily to the result 
of Kim's prescription in our
single-$W$ case. The SV (more precisely, the singularity function) is:
\begin{equation}
\Sigma({\cal M},\vec l,\vec p_{_T})={l_{_T}^2+2\,l_3^2\over 4\, l_{_T}^4}\;
\Sigma_T({\cal M},\vec l_{_T},\vec p_{_T})
\label{SV}
\end{equation}
with $\Sigma_T$ as in Eq.~(\ref{calMeq2}), and $M$ substituted for
${\cal M}$, as its role will now be that of a trial mass.
For $\vec p_{_T}=0$ this SV reduces to:
\begin{equation}
\Sigma_0({\cal M},\vec l,\vec p_{_T})={l_{_T}^2+2\,l_3^2\over 4\, l_{_T}^4}\;
{\cal M}^2\;({\cal M}^2-4\,l_{_T}^2)
\label{SVbis}
\end{equation}

Refer for a moment to the limit $\Gamma\to 0$ for the $W$ width and
a situation with no measurement uncertainties.
Consider a set of $N$ real or MC generated events, i.e.~a list of values 
of $(\vec l,\vec p_{_T})$ and the histograms $dN({\cal M})/d\sigma$
of the corresponding values of $\sigma=\Sigma({\cal M},\vec l,\vec p_{_T})$, 
for different choices of $\cal M$. For ${\cal M}=M_W$, 
the real or ``MC true" value of the $W$ boson mass, the singularity
is at $\sigma=0$, $dN({\cal M})/d\sigma$ peaks at that point and
vanishes for $\sigma<0$. For  a fixed data set  
and varying ${\cal M}$, the function $dN({\cal M})/d\sigma$ varies in shape,
but obviously not in statistically useful content. We shall later
illustrate these points in detail.

The use of an  ``implicit" variable ${\cal{M}}$ may seem to be an
overkill. In the single-$W$ case with $\vec p_T=0$, it is. 
One could equally well erase ${\cal M}$ in Eq.~(\ref{SVbis})
and use the SV:
\begin{equation}
\Sigma_l({\cal M},l)={l_{_T}^2+2\,l_3^2\over l_{_T}^2}\, ,
\label{SVtris}
\end{equation}
which, in conjunction with ${\cal M}^2=4\,l_T^2$, embodies
two projections of the full distribution $dN/(dl_T\,dl_3)$.

Contrariwise, one could 
make the singularity condition into a singularity variable
with an implicit  ${\cal M}$:
\begin{equation}
\Sigma_T({\cal M},l_{_T})\equiv {\cal M}^2-4\,l_{_T}^2
\label{SVtetra}
\end{equation}
and consider the distributions $dN({\cal M)}/d\sigma_{_{T}}$. 
But the information that these distributions
contain is precisely the same as
that of the distribution $dN/dl_{_T}^2$, the corresponding histograms are just
mirror reflected and shifted relative to one another.

The above unfavorable commentaries on implicit variables are by
no means general. Even in the single-$W$ case, for $\vec p_T\neq 0$,
it will not be possible to ``erase" $\cal M$ from Eq.~(\ref{SV})
in the same cavalier spirit in which we erased it from Eq.~(\ref{SVbis})
to obtain Eq.~(\ref{SVtris}). Singularity variables should
 be of particular practical relevance in problems with more than
one unknown mass or unobservable particle, for which the labor
of making templates for all possibilities may be out of the question. 
There, at least at the discovery stage, ``clever" variables may be useful 
to zoom kinematically to the relevant mass ranges before a full analysis 
is to be contemplated, as discussed in \cite{CR}.

\section{The quest for an optimal variable}
\label{sec:QOV}

It is instructive to consider a trivial example with one visible variable,
$l$, and a single invisible one, $x$, constrained by the ``Euclidean phase space" equation
\begin{equation}
\Phi := x^2+l^2-M^2=0
\label{trivial}
\end{equation}
This apparently arbitrary instance actually corresponds to an imaginable
process, that of a particle decaying into an invisible one, $X$, and a 
visible one that happens to be at rest. The longitudinal momentum 
of $X$ is $x$ and its transverse one, $l$, is measured via the usual
transverse balance. $M$ is a combination of the masses involved \cite{BP}.

The value of the unknown quantity $M$ in Eq.~(\ref{trivial})
is encoded in the $l$-distribution.
 The Jacobian matrix is $D=\partial \Phi/\partial x=2 x$. 
The constraint that its rank be reduced is $x=0$, resulting in the SCs
$l=\pm M$. For a given ``observed" $l$, there are two points $P$ in
$\Phi$. Their nearest singularity is the point $S$,
as illustrated in Fig.~\ref{fig:phasespacetrivial}.

\begin{figure}[htbp]
\begin{center}
\includegraphics[width=0.4\textwidth]{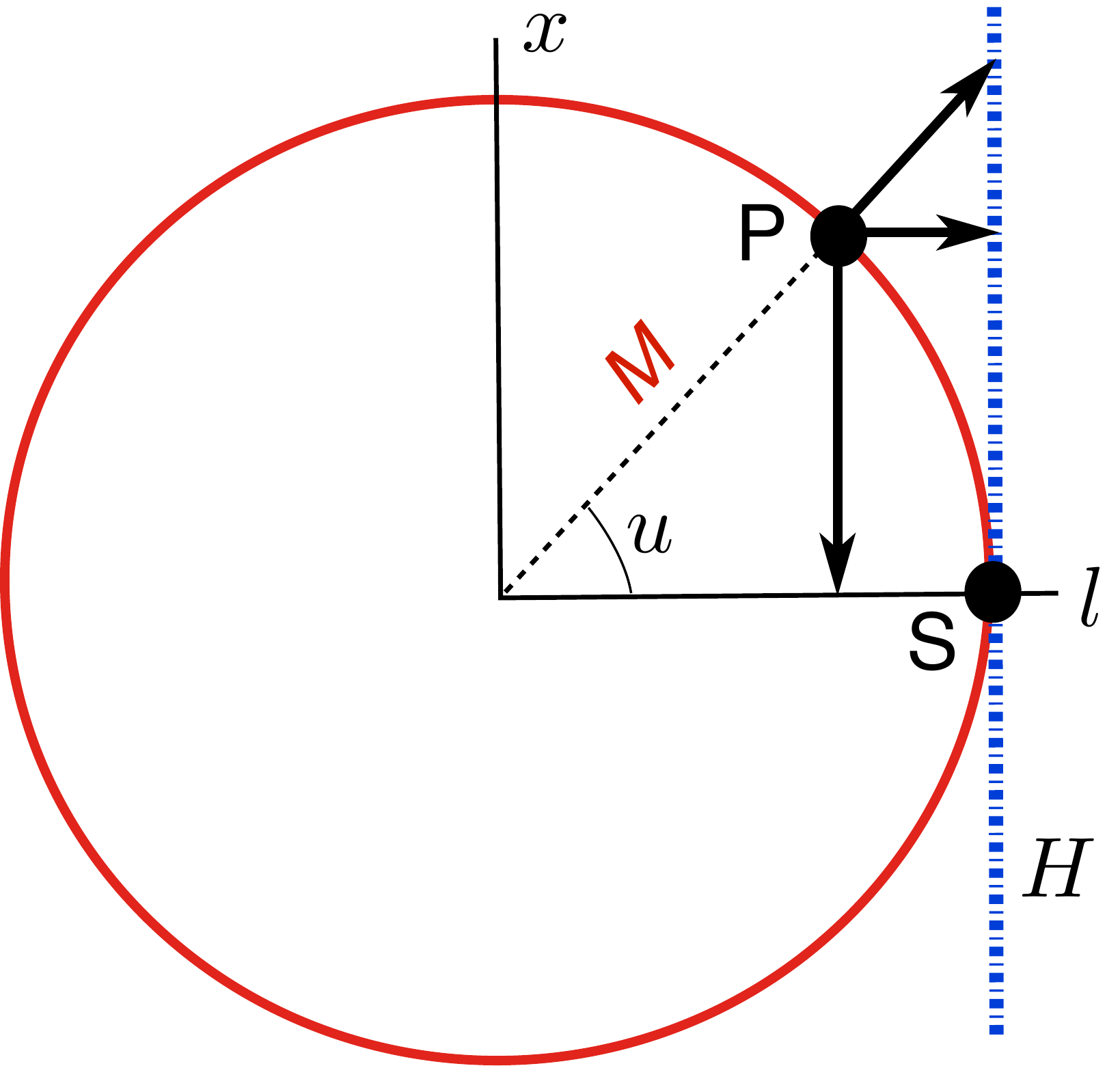}\\
\caption{$P$ is a point in ``phase space" of which only the corresponding
$l$ is measured. $S$ is the closest singularity to it.
The length of the three arrows and the angle $u$ are used to construct
various singularity variables.
\label{fig:phasespacetrivial}}
\end{center}
\end{figure}

Following Kim's method \cite{Kim}, we obtain for the SV
\begin{equation}
\Sigma_K({\cal M},l)=u^2\equiv\left[\arccos {|l|\over {\cal M}}\right]^2,
\label{SigmaKim}
\end{equation}
 proportional to the squared (angular or geodesic) $P$ to $S$ 
distance measured on the $\Phi$ surface.
In a less trivial case, the resulting SV would have been 
the same distance on the quadratic approximation to 
$\Phi$ around $S$.

There is nothing sacred about the elegant result of Eq.~(\ref{SigmaKim}).
There are other SVs that (up to an overall normalization)
coincide with $u$ to second order. Three examples, illustrated in
Fig.~\ref{fig:phasespacetrivial}, are: 
\begin{itemize}
\item{(1)} The distance between $P$ and
the hyperplane, ${ H}$, tangent to $\Phi$ at $S$ (the dotted vertical line, in this case).
This distance is the horizontal arrow.
\item{(2)}  The $P$ to ${ H}$ distance along the
normal direction to $\Phi$ at $P$: the slanted arrow.
\item{(3)} The square of the length of the vertical arrow.
\end{itemize} 
In the notation of Eq.~(\ref{SigmaKim}) and normalized so that they
coincide with $\Sigma_K$ to ${\cal O}(u^2)$, these SVs are:
\begin{eqnarray}
\Sigma_1({\cal M},l)&=& 2\,\left[1-\cos u\right]\label{Sigma1}\\
\Sigma_2({\cal M},l)&=& 2\,\left[1/\cos u-1\right]\label{Sigma2}\\
\Sigma_3({\cal M},l)&=& \sin^2 u\label{Sigma3}
\end{eqnarray}
Note that $\Sigma_1$ is the 2D analog of the singularity
condition used as a SV, as in Eq.~(\ref{SVtetra}). That is
to say, it is equivalent to the transverse mass distribution.

Is any of these SVs in Eqs.~(\ref{SigmaKim}) to (\ref{Sigma3})
 ``the best" in some useful sense?
To answer, consider the distributions of the
numerical values $\sigma$ of the various $\Sigma_i$
functions, for fixed $M$ (a zero width resonance):
\begin{eqnarray}
&&{\cal H}_i(\sigma,M,{\cal M})  \equiv {dN\over d\sigma}\equiv
\nonumber\\
&&\int dx\,dl\,
\delta(x^2+l^2-M^2)\,\delta[\sigma-\Sigma_i({\cal M},l)]
\label{sigmaDist}
\end{eqnarray}
Recalling Eq.~(\ref{trivial}), and in particle physics language, 
$dx\,dl\,\delta(\Phi)$ is the phase space, ${\cal H}_i$
is the distribution of the $\Sigma_i$ values.
 Monte Carlo generated ``diagonal" histograms, 
${\cal H}_i(\sigma,{M},{ M})$, would be  the templates
for various trial choices of ${ M}$.

In the four cases of Eqs.~(\ref{SigmaKim}) to (\ref{Sigma3}),
with the notation $\rho\equiv {\cal M}/M$, and normalized to
unit integral in the allowed range of the corresponding $\sigma$,
the distributions are
\begin{eqnarray}
&&\!\!{\cal H}_K={\rho \sin\sqrt\sigma\over \pi\sqrt{1-\rho^2\cos^2\sqrt\sigma}},
\;\sigma \in [\arccos^2 \rho^{-1},\pi^2/4]\nonumber\\
&&\!\!{\cal H}_1={\rho \over \pi\sqrt{1-\rho^2+\rho^2(\sigma-\sigma^2/4)}},
\;\sigma \in [2(1-\rho^{-1}),2]\nonumber\\
&&\!\!{\cal H}_2={4\rho\over \pi(2+\sigma)\sqrt{(2+\sigma)^2-4\rho^2}},
\;\sigma \in [2(\rho-1),\infty)\nonumber\\
&&\!\!{\cal H}_3={\rho \over \pi\sqrt{1-\rho^2(1-\sigma)}\sqrt{1-\sigma}},
\;\sigma \in [1-\rho^{-2},1]
\label{Hs}
\end{eqnarray}
In the simple case at hand, one need not refer
 to ``nondiagonal" histograms ${\cal H}_i(\sigma,{M},{\cal M})$,
that involve the implicit variable ${\cal M}\neq M$.
In more blind searches with several unknown masses this
may no longer be the case. Moreover the nondiagonal
histograms provide one way to ascertain the ``goodness"
of their SV.

To quantify the amount by which the distribution of a given SV is 
sensitive to the difference between a ``true" mass ${\cal M}=M$ and a variation
thereof, ${\cal M}=M+\Delta M$, define the ``statistical squared derivative", $\hat\chi^2$,
and its integral \cite{mmm}
\begin{eqnarray}
\hat\chi_i^2(\sigma)&\equiv& {1\over {\cal H}_i(\sigma,M,M)}\;
\left[{\partial {\cal H}_i(\sigma,M,{\cal M})\over \partial {\cal M}}
\right]^2_{{\cal M}=M}\nonumber\\
D_i&=&\int_{\sigma_{\rm min}}^{\sigma_{\rm max}}\hat\chi^2_i(\sigma)\,d\sigma 
\label{SD}
\end{eqnarray}
The notation reflects the parentage of $\hat\chi^2$ with the usual $\chi^2$
measure; it is also the square of the geometrical mean between ordinary and logarithmic
derivatives. ``Statistical" reflects the fact that $\hat\chi^2(\sigma)$ is
a local measure of a variation relative to the one expected from a standard
deviation of 1$\sigma$ size. 
In this hypothetical case
with sharply defined cuts in $\sigma$,  $\hat\chi^2$ is
singular at $\sigma=0$. Regularizing the singularity with a cut $\sigma>\sigma_0>0$
we obtain:
\begin{equation}
\begin{split}
& D_K\underset{\sigma_0\downarrow 0}{\sim}
{2\over 3\pi}\sigma_0^{-3/2}\left(1+2\,\sigma_0\right) +o(1),
\\
&D_1\underset{\sigma_0\downarrow 0}{\sim}
{2\over 3\pi}\sigma_0^{-3/2}\left(1+{15\over 8}\sigma_0\right) +o(1),
\\
&D_2\underset{\sigma_0\downarrow 0}{\sim}
{2\over 3\pi}\sigma_0^{-3/2}\left(1+{21\over 8}\sigma_0\right) +o(1),
\\
&D_3\underset{\sigma_0\downarrow 0}{\sim}
{2\over 3\pi}\sigma_0^{-3/2}\left(1+{3\over 2}\sigma_0\right) +o(1).
\end{split}
\label{derivatives}
\end{equation}

The singularities
of the different $H_i$ are all $\propto 1/\sqrt \sigma$ and have been equally 
normalized by construction (and for a fair comparison). The sensitivity to
the value of $M$ is maximal close to the singularity. This sensitivity puts
the SVs  of Eqs.~(\ref{SigmaKim}) to (\ref{Sigma3}) in the ``goodness" order
\begin{equation}
\Sigma_2\succ \Sigma_K\succ \Sigma_1\succ \Sigma_3
\label{order}
\end{equation}
dictated by the second term in brackets in Eqs.~(\ref{derivatives}).
The fully ``orthogonal" SV $\Sigma_2$ is the contest's winner.
The usual transverse mass distribution ($\Sigma_1$ in this
simplification) does not fare well.

So far there seems to be no compelling reason not to have made the above 
variable-comparing analysis with $M=\cal M$ for starters. But in a more realistic case
$M$ would stand for the central value of a distribution of non zero natural width,
while $\cal M$ is just an auxiliary quantity introduced for analysis purposes.

To illustrate the above, and to
convey the numerical meaning of Eqs.~(\ref{derivatives}), substitute the sharp
definition of $M$ in Eqs.~(\ref{trivial},\ref{sigmaDist}) by the one corresponding
to a resonance of mass $M$ and width $\Gamma$:
\begin{equation}
\delta(x^2+l^2-M^2) \to {1\over \pi}\, \frac{ M\, \Gamma }{ 
   \left(l^2+x^2-M^2\right)^2+M^2\,\Gamma^2}
\label{resonance}   
\end{equation}
This corresponds to ``spreading" the circle of Fig.~(\ref{fig:phasespacetrivial})
and ``scanning" it with circles of varying --but sharply defined--  ${\cal M}$,
with the help of different ``$\Sigma$" scanners.

Results for the distributions for Kim's variable and the 
orthogonal SV are shown in the upper Fig.~(\ref{fig:SigmaDistribs}).
The lower figure shows their $\hat\chi_i^2(\sigma)$ around the
$\sigma=0$ singular point, the domain to which the ${\cal H}_i$ distributions
are most sensitive to the unknown ${\cal M}$. The figures are drawn for
$M={\cal M}=1$, $\Gamma=0.3$, showing how the orthogonal $\Sigma_2$
is better than $\Sigma_K$. However, the difference is not large and, for a narrow 
resonance (or one whose width is masked by detector effects) it would be
negligible, as the relative differences close to $\sigma=0$ between the $\hat\chi_i^2(\sigma)$ 
of the various SVs diminish linearly as $\Gamma/M\to 0$.

   \begin{figure}[htbp]
\begin{center}
\includegraphics[width=0.4\textwidth]{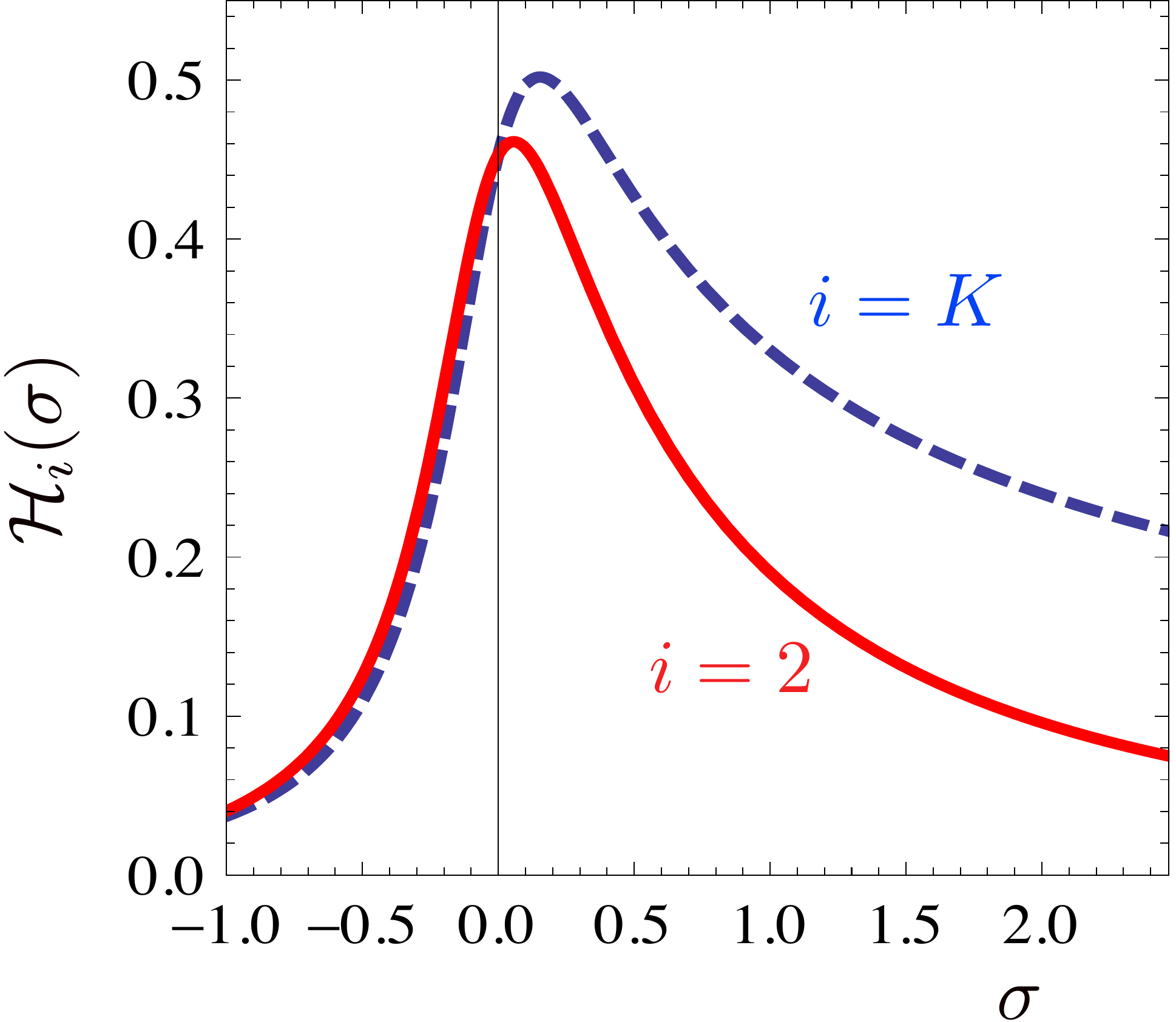}\\
\vspace{.5cm}
\includegraphics[width=0.4\textwidth]{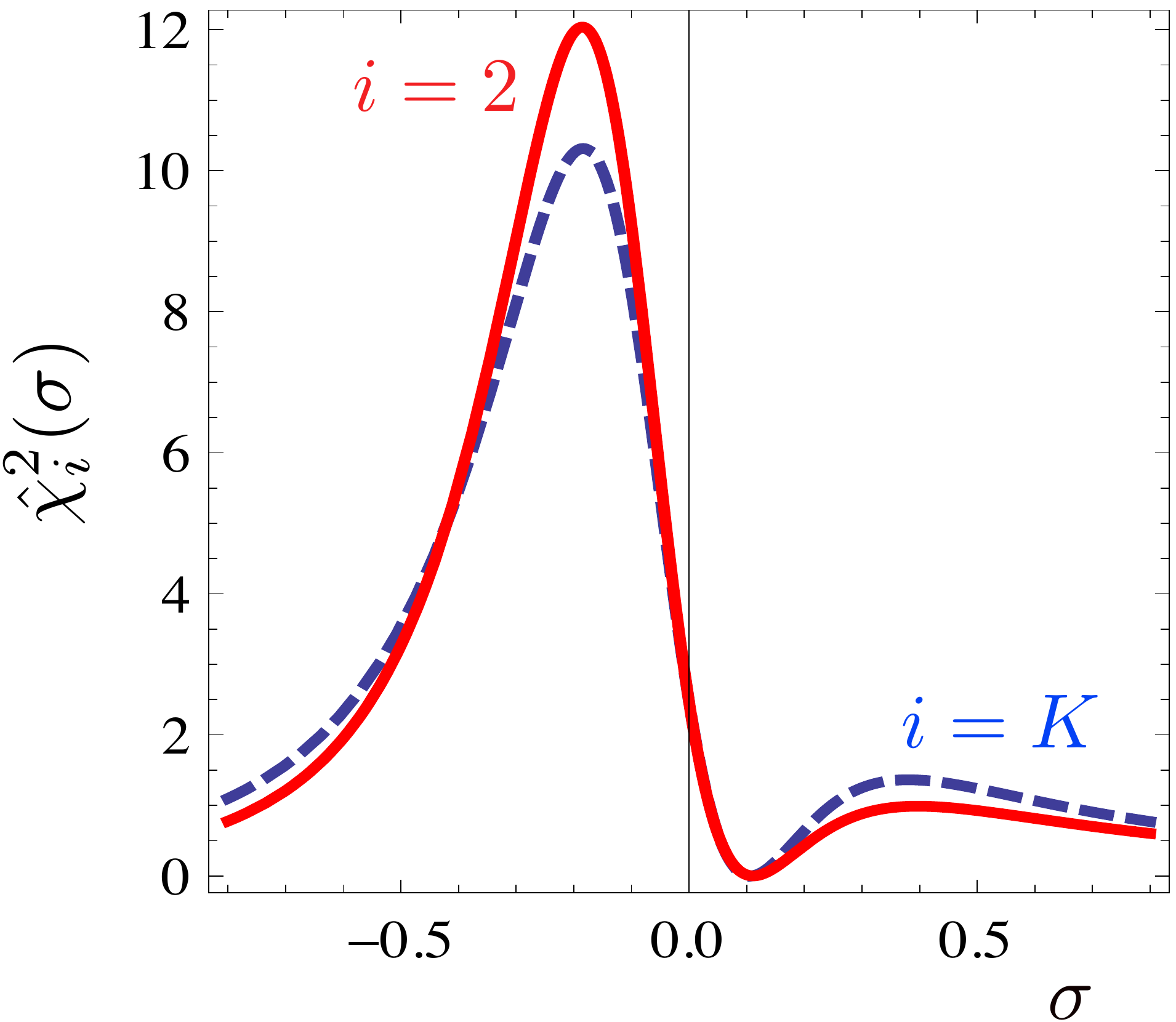}\\
\caption{
Top: the $dH_i(\sigma,M,\Gamma,{\cal M})/d\sigma$ distributions for
the SVs $\Sigma_i$, $i=K,2$ for $M={\cal M}=1$, $\Gamma=0.3$.
Bottom: the corresponding statistical squared derivatives.
  \label{fig:SigmaDistribs}}
\end{center}
\end{figure}

The $D_i$ integrals of Eq.~(\ref{SD}) over their
complete respective kinematical domains are numerically
similar, apparently demonstrating that, {\it in toto,} all variables
are statistically equivalent. In practice this is not the case. The signal-to-noise
ratios of the distributions are increasingly unfavorable as
one moves away from the $\sigma_i\sim 0$ neighborhood of
the signal's peak.

We have proven that $\Sigma_2$ is better than others, but not that it
is the best. Its optimality, however, appears to be intuitively obvious.
The phase space $\Phi$ of Eq.~(\ref{trivial}) simply scales as $M$
changes. The optimal SV ought to maximize the dependence on $M$
at every point in phase space. This dependence is maximal in the
direction orthogonal to $\Phi$. The variable $\Sigma_2$ measures
a distance to the nearest singularity, in that preferred direction.

\section{Induced singularities}
\label{sec:IS}

Let us return to the case of single-$W$ production and model the simplified
$\vec p_T=0$ instance as stated in the ending paragraph of \S \ref{sec:I}, that is, to leading order.
We use the quark and antiquark parton distribution functions of \cite{Durham} at an LHC
energy of $\sqrt{s}=7$ TeV and apply the cuts $|l_T|>10$ GeV and
$|\bar \eta |< 2.3$ to the charged lepton. We ignore the difference between $W^+$ and $W^-$
production.

We choose to present results for the distribution of the values, $\sigma$, of the function:
\begin{equation}
\Sigma({\cal M},l)=({l_{_T}^2+2\,l_3^2})\;
{\cal M}^2\;({\cal M}^2-4\,l_{_T}^2),
\label{SVpenta}
\end{equation}
which differs from Eq.~(\ref{SVbis}) by a factor $4\,l_T^4$. This does not affect the
arguments to follow. Moreover, in conjunction
with the transverse mass ($4\,l_T^2$) distribution, the use of Eqs.~(\ref{SVbis}) or
(\ref{SVpenta}) are equivalent.

A heedless use of Eq.~(\ref{SVpenta}) results in an interesting surprise, illustrated
in the top panel of Fig.~\ref{fig:KimDist}. The histogram has two peaks, one of them
significantly above the expected singularity at $\sigma=0$. The peaks fuse as one lets
the $W$ have its rather narrow width, $\Gamma/M\simeq 0.02$, as illustrated in the
lower panel of Fig.~\ref{fig:KimDist}. Still, the fused peak is not just the expected
singularity at the origin of the SV and the issue calls for understanding.

\begin{figure}[htbp]
\begin{center}
\includegraphics[width=0.4\textwidth]{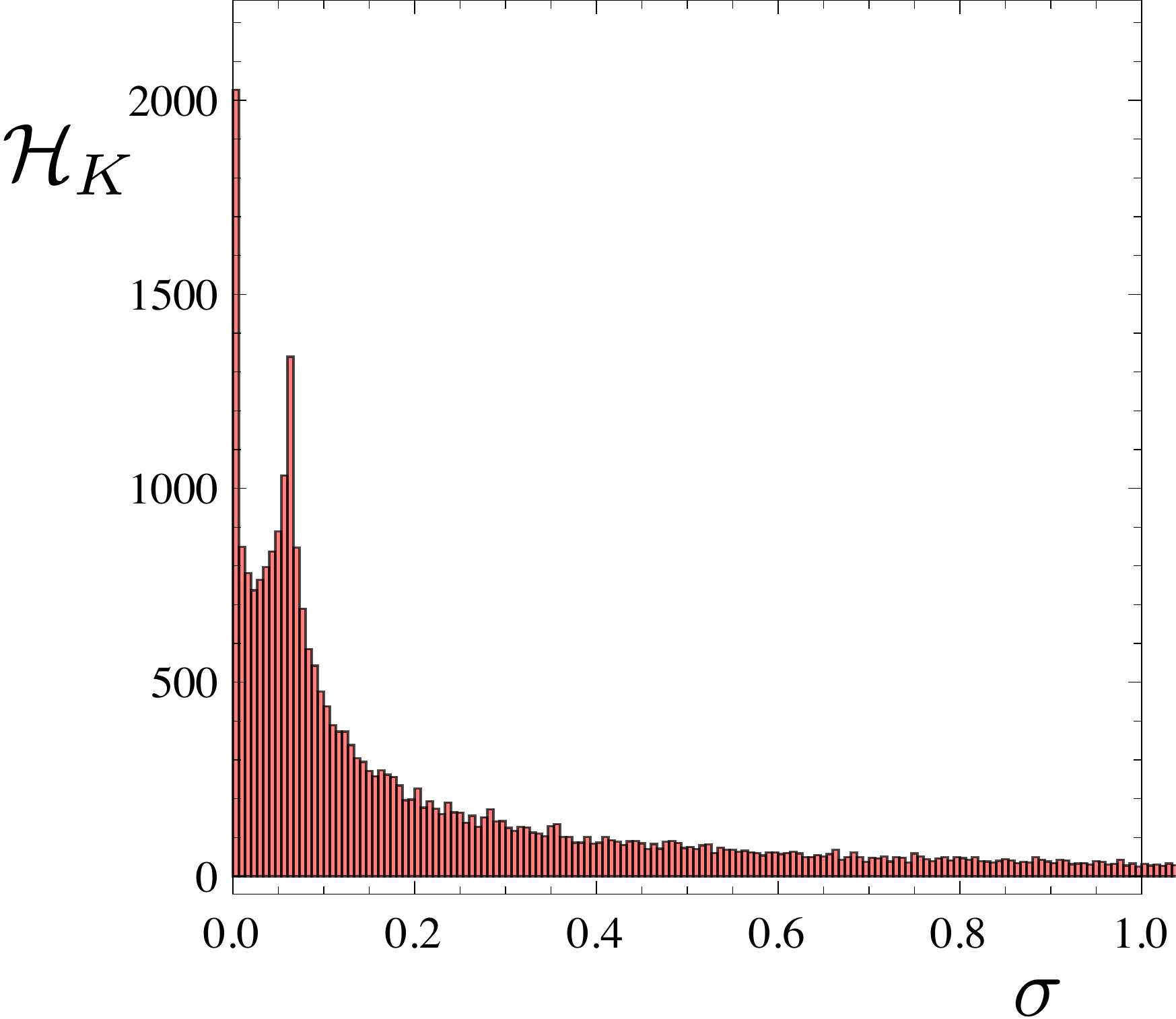}\\
\vspace{.5cm}
\includegraphics[width=0.4\textwidth]{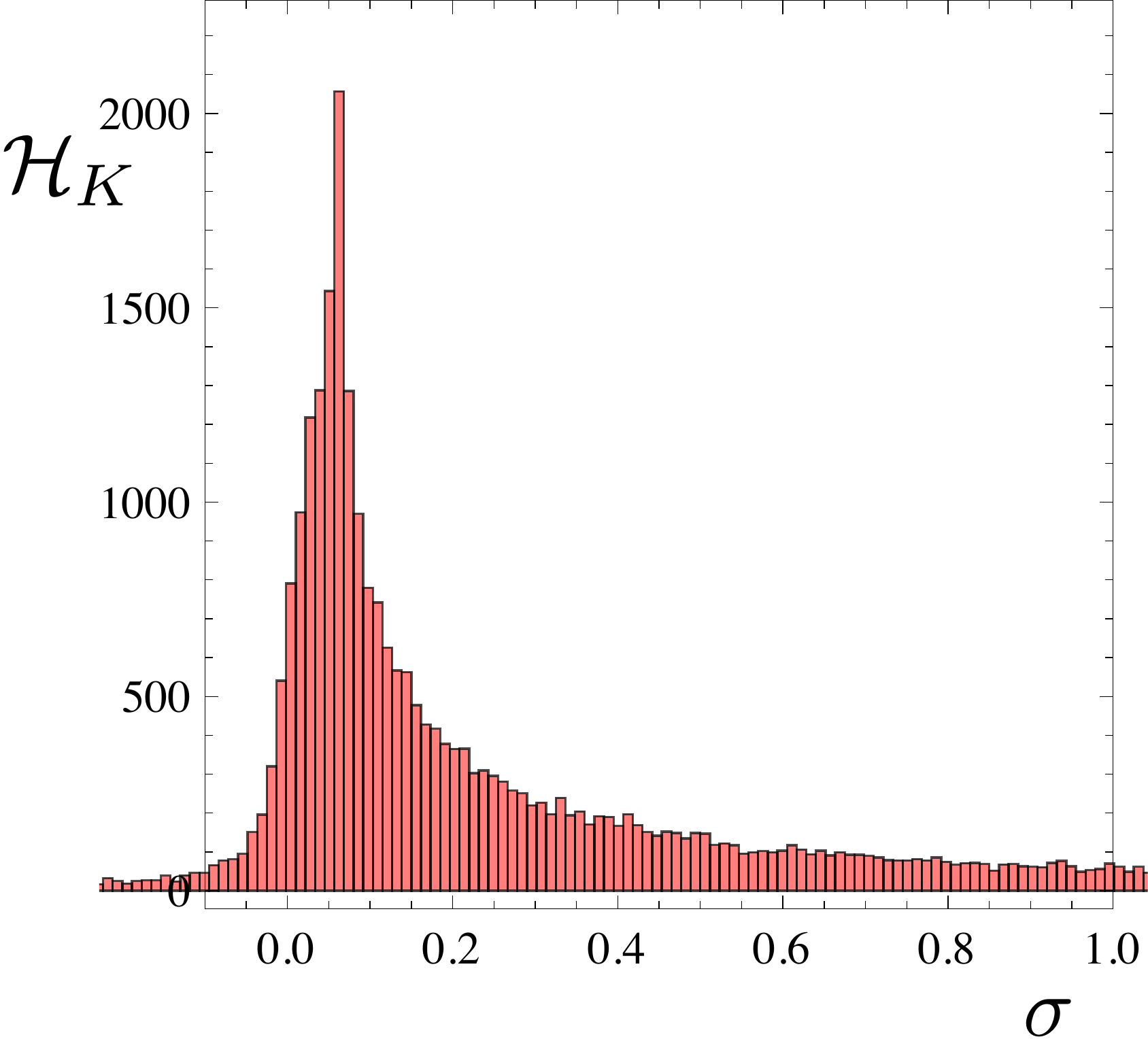}\\
\caption{Top: The singularity variable of Eq.~(\ref{SVtris}) results,
for a narrow resonance, in a distribution with an extra singularity
away from $\sigma=0$.
Bottom: The small width of the $W$ suffices to merge the singularities,
shifting the resulting peak away from $\sigma=0$.
  \label{fig:KimDist}}
\end{center}
\end{figure}

Consider restricting the phase space of Eqs.~(\ref{PhaseSp0}) and 
Fig.~\ref{fig:phasespace} to its slices at fixed longitudinal momentum
 of the $W$, $W_3=x_3+l_3$, shown in these plots as (green) ellipses
(in practice this can only be done at a monochromatic $e\nu_e$ collider).
The distribution ${\cal H}(\sigma,M,{\cal M},W_3)$ is shown on the
upper Fig.~\ref{fig:W3Dist}, for $M={\cal M}=1$, $W_3=2$. It
has two singularities besides the one expected at $\sigma=0$.

\begin{figure}[htbp]
\begin{center}
\includegraphics[width=0.4\textwidth]{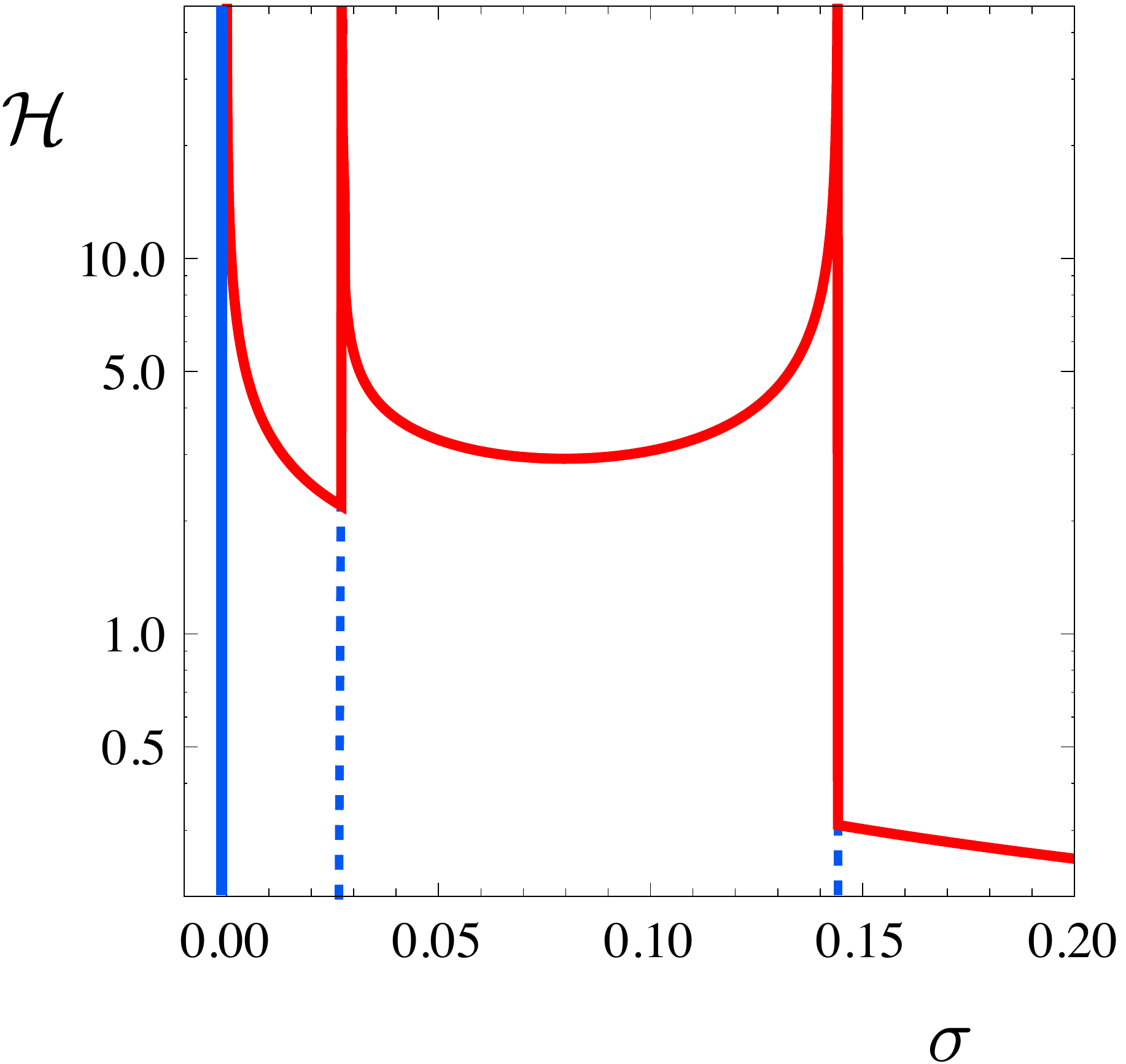}\\
\vspace{.5cm}
\includegraphics[width=0.4\textwidth]{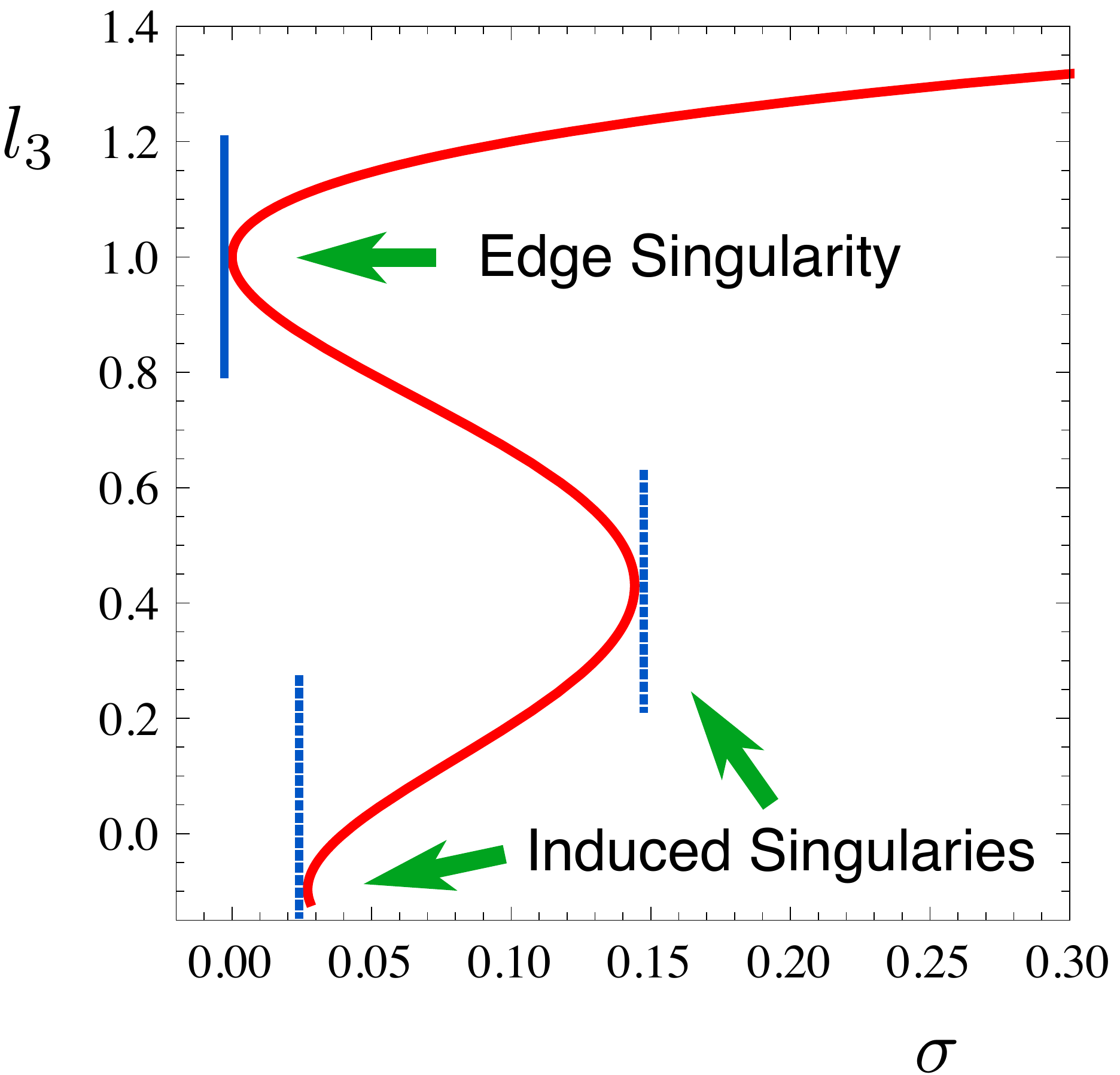}\\
\caption{
Top: The phase space of Eqs.~(\ref{PhaseSp0}) and  Fig.~\ref{fig:phasespace}
for a fixed $W_3=x_3+l_3=2$ results, for a narrow resonance, in a triple peaked
distribution (all quantities in units of $M=1$ units). The singularities occur
at values of $\sigma$ where the phase space $\Phi(l_3,\sigma)$
has vertical $l_3$ projections.
  \label{fig:W3Dist}}
\end{center}
\end{figure}

The origin of the singularities is clarified in the lower Fig.~\ref{fig:W3Dist},
where the curve is the phase space $\Phi(l_3,\sigma)$, again 
for $M=1$, $W_3=2$. A uniform distribution of events along
$\Phi(l_3,\sigma)$, projected on the $\sigma$ axis, has three cumulation
points at the projections of the vertical tangents. The one at the edge
is the expected $\sigma=0$ singularity, the other two are {\it induced
singularities}. In these $M_W=1$ units, for $W_3<1$ there is no induced
singularity, for $W_3=1$ there is one and for $W_3>1$ there are two.
One induced singularity survives the integration over the $W_3$
distribution, as shown in Fig.~\ref{fig:KimDist}. 

The source of the induced 
singularities is the specific form of the SV in Eq.~(\ref{SVpenta}) --or of the 
formal SV of Eq.~(\ref{SVbis})-- which results in a  fixed-$W_3$
phase space the curvature of whose surface is not everywhere of the same sign.
The induced singularities are not endpoints, but are event accumulation points
for the same reason as the endpoints, i.e.~the tangent manifold to the phase space
at their locations contains invisible directions.

In a process with just one mass scale to disentangle, the complications
we just discussed are a lesser problem. In a process with more than
one mass scale, they are a putative source of confusion. The fully
orthogonal SV $\Sigma_2$ of Eq.~(\ref{Sigma2}) does not result in
induced singularities. 

\section{Results}
\label{sec:R}

For the single-$W$ case at hand, consider the ``fully orthogonal" variable
akin to $\Sigma_2$ in Eq.~(\ref{Sigma2}). We call it $\Sigma_A$ and discuss
it first in the $\vec p_T=0$ instance. Its geometrical interpretation is depicted
in Fig.~(\ref{fig:phasespace2}); $\Sigma_A$ is a measure of the length of the 
arrow, which is orthogonal to a phase space point $P$ with coordinates
$(l_T,l_3,x_3)$ and ends in the plane tangent to the phase space surface at
the singularity line.

\begin{figure}[htbp]
\begin{center}
\includegraphics[width=0.43\textwidth]{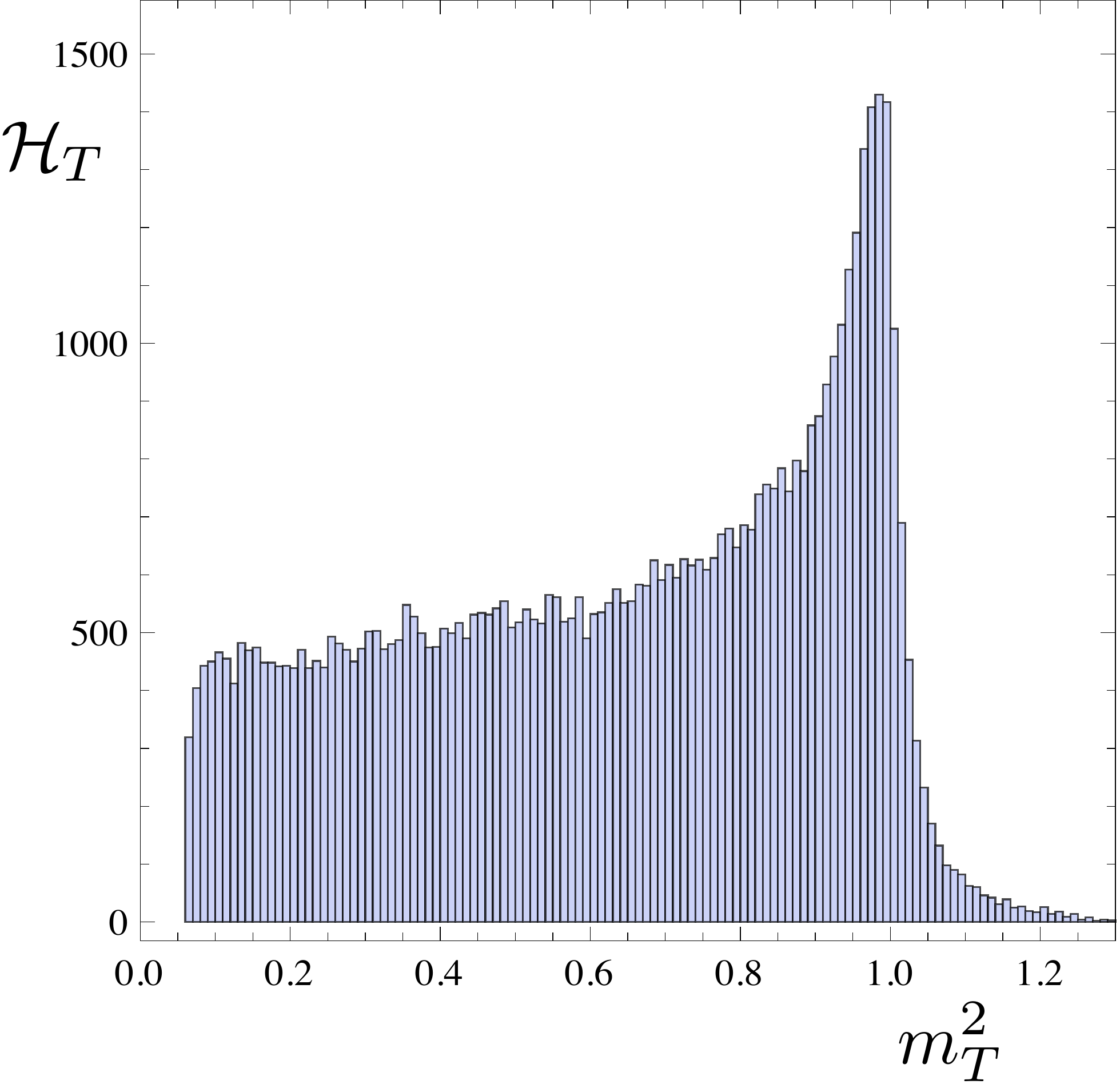}\\
\vspace{.3cm}
\includegraphics[width=0.45\textwidth]{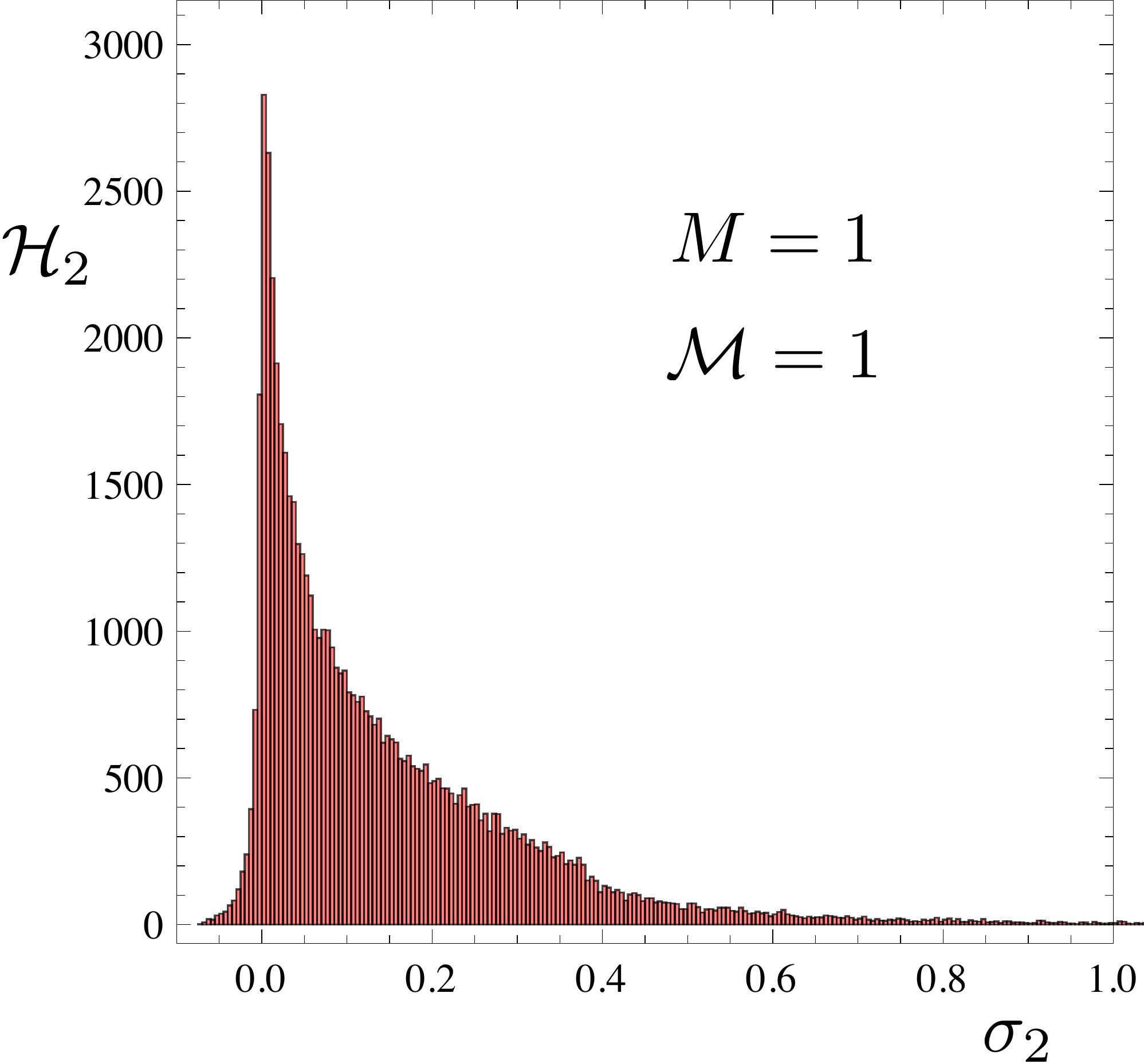}\\
\vspace{.2cm}
\includegraphics[width=0.43\textwidth]{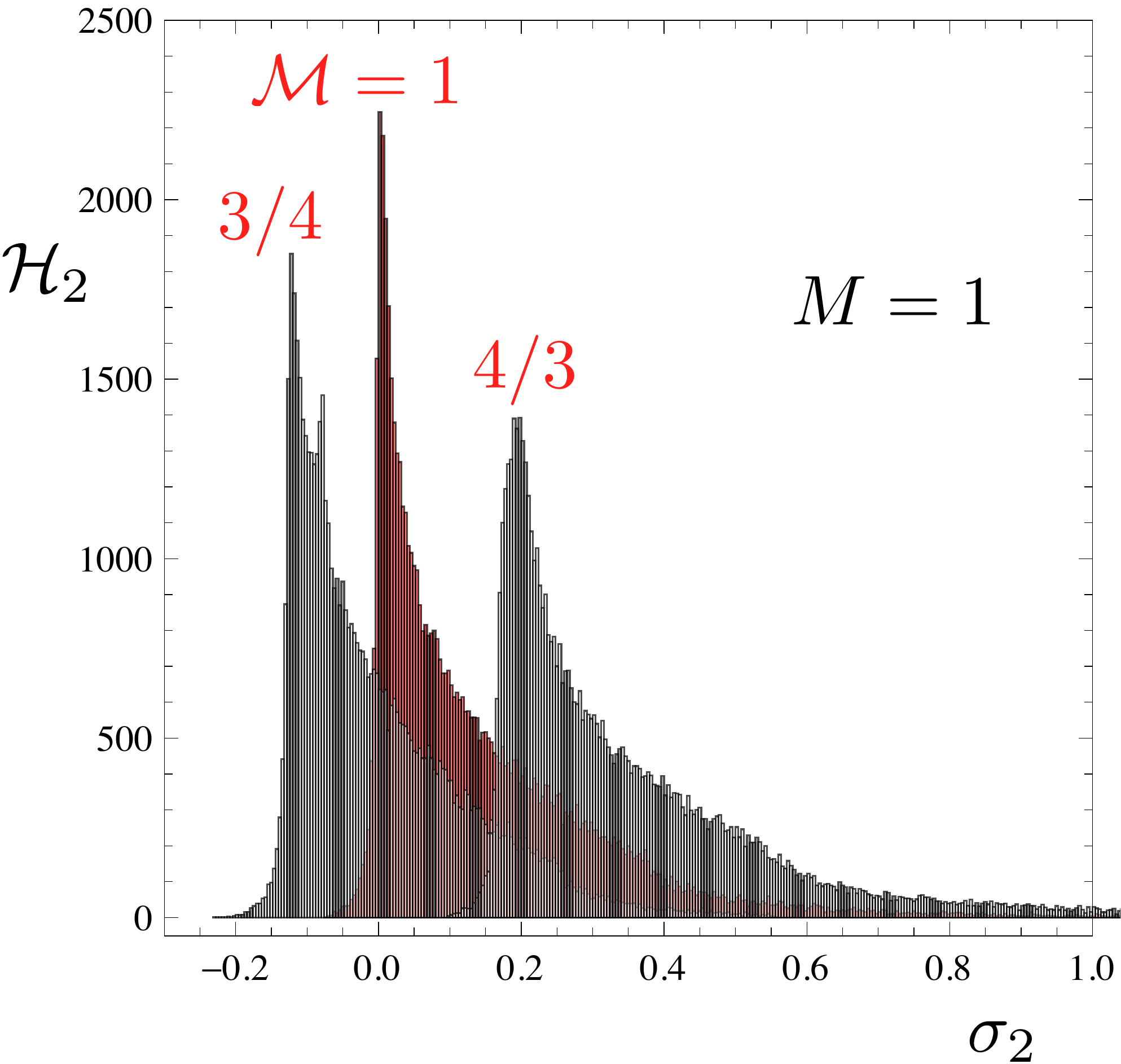}\\
\caption{Top: Histogram ${\cal H}_T$  of the distribution of the
square of the transverse mass, for $M=1$. Center: Histogram ${\cal H}_2$
of the distribution of the values $\sigma_2$ of the optimal SV
$\Sigma_A$ of Eq.~(\ref{SigmaA}), for ${\cal M}=M=1$. Bottom: same as center,
for different values of $\cal M$. In all cases  ${\vec p}_T=0$.
  \label{fig:SingleWSingCond}}
\end{center}
\end{figure}

Define the unit vector $\vec n$ orthogonal to the surface
 $\Phi(l_T,l_3,x_3,{\cal M})$ of Eq.~(\ref{PhaseSp0}):

\begin{eqnarray}
\vec N&\equiv&(N_1,N_2,N_3)=\left({\partial \Phi/ \partial l_T},{\partial \Phi/ \partial l_3},
	{\partial \Phi/ \partial x_3}\right)\nonumber\\
\vec n&=&\vec N/ |N|
\label{n}
\end{eqnarray}
The length, $\Sigma_A$, of the orthogonal segment joining $P$ with a point
in the plane tangent to the singularity is such that 
\begin{equation}
\Sigma_A\;\mid\;{{\cal M}\over 2}=l_T-\Sigma_A\,n_1
\end{equation}
More explicitly  
\begin{eqnarray}
\Sigma_A(l_T,l_3,{\cal M})&=&\frac{{\cal M}/2-l_T}  {{2 \, l_T \left({\cal M}^2+W_3^2\right)}}\times
\nonumber\\
&&\!\!\!\!\!\!\!\!\!\!\!\!\!\!\!\!\!\!\!\!\!\!\!\!\!\!\!\!\!\!\!\!\!\!\!\!\!\!\!\!\!\!
\sqrt{{\cal M}^4 \left( 2
   l_3^2+W_3^2-2 l_3 W_3\right)+8
   W_3^2 l_T^4
   \atop  +4 \,l_T^2\,
   \left({\cal M}^4+{\cal M}^2 \,W_3^2+W_3^4\right)}
\nonumber\\ 
\nonumber\\ 
W_3&\equiv& l_3+x_3(l_T,l_3,{\cal M})
 \label{SigmaA}
\end{eqnarray}
with $x_3$ as in Eq.~(\ref{x3solution}). For each $(l_T,l_3)$ pair
(an event) there are two equal probability solutions, the two
roots of the equation. In generating events we chose at
random the $\pm$ sign in Eq.~(\ref{x3solution}). 

We show in Fig.~(\ref{fig:SingleWSingCond}) the $\vec p_T=0$ results for 
the $m_T^2$ and $\Sigma_A$ distributions. All three graphs
are generated for a peak mass of the $W$, $M=1$. As shown in the bottom
figure, for a trial mass ${\cal M}\neq M$ the peak of the distribution
shifts away from $\sigma_A=0$, becoming wider and, for ${\cal M}<M$,
double peaked: there is for this ``bad" choice an induced singularity, even for
the optimal SV. Naturally, the histograms with ${\cal M}\neq M$
are not statistically independent from the ${\cal M}= M$ one. 
While they may be used to ``focus" on the correct choice of ${\cal M}$,
the extraction of information on the $W$ boson mass would ultimately
hinge on a set of templates for ${\cal M}=M$ values close to its
currently measured value.

The value of $x_3$ is not always real.
When the value of $l_T^2$ chosen by the Lorentzian distribution of physical (or MC
generated) values of $M_W$ is such that $4\,l_T^2 > {\cal M}^2$, $x_3$ involves
the square root of a negative number. There is nothing pathological about these
events. The way to ``recover" them is to set:
\begin{equation}
{\rm If} \;{\rm Im}\left(\Sigma_A\right)\neq 0;\;\;{\rm then}
\;\; \Sigma_A\to - {\rm Abs}(\Sigma_A)
\end{equation}
In the middle Fig.~(\ref{fig:SingleWSingCond}), for example, the recovered
events are those at $\sigma_2<0$.

\section{Correlations}
\label{sec:C}

It is clear that the transverse mass --or its equivalent $\Sigma_T$
of Eq.~(\ref{SVtetra})-- and the SV of Eq.~(\ref{SigmaA}) are
highly correlated. They both vanish at the singularity as ${\cal M}-2\,l_T$.
To illustrate the point, define the variable
\begin{equation}
\Sigma_t={\cal M}-2\,l_T
\label{Sigmat}
\end{equation}
which has the same mass dimensionality as $\Sigma_A$ and, close
to the singularity, carries the same information as $\Sigma_T$.
The double histogram $dN/d\Sigma_A\,d\Sigma_t$, shown in
Fig.~\ref{fig:correlations}, illustrates the expected correlation.

\begin{figure}[htbp]
\begin{center}
\hspace{-1.5cm}
\includegraphics[width=0.55\textwidth]{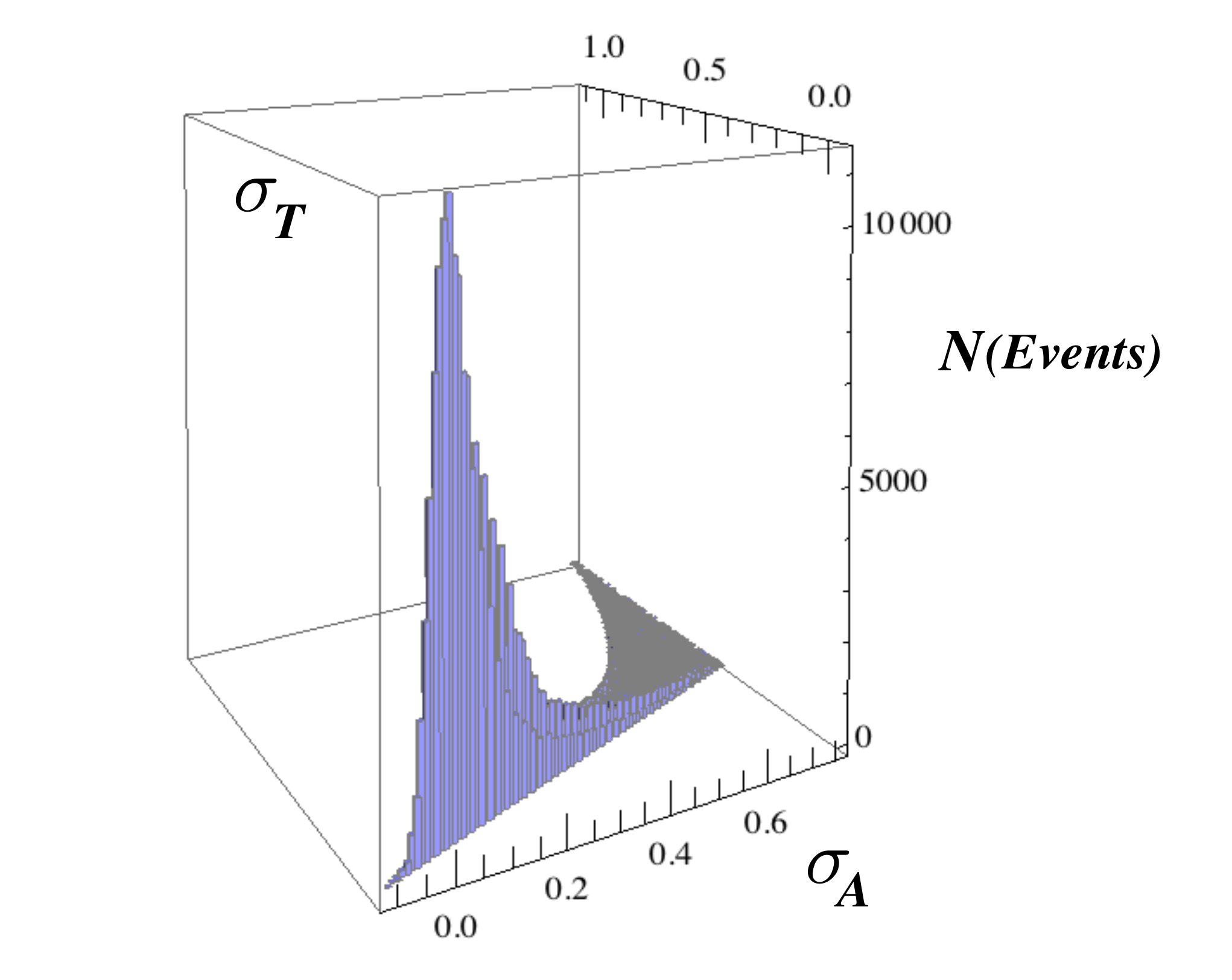}
\caption{The correlation between the SV of Eq.~(\ref{SigmaA})
and the SC expressed as the SV of Eq.~(\ref{Sigmat}), for
${\cal M}=M=1$.
  \label{fig:correlations}}
\end{center}
\end{figure}

Naturally,  correlations between observables constitute
a weakness of their ensemble, to which we shall come back in the
conclusions. Suffice it to say here that in the ``signal only" case at hand, 
there is only one mass scale to extract from the data: the correlations
are unavoidable.

\section{The general case}
\label{sec:GC}

In Figs.~(\ref{fig:phasespace},\ref{fig:phasespace2}) we have profited
from the fact that the $p_T=0$ phase space of Eq.~(\ref{PhaseSp2})
is a function of $l_T^2$ to plot the phase space for negative and
positive $l_T$. For $p_T\neq 0$ this is no longer possible. Let $l_T$
and $p_T$ be the moduli of the corresponding vectors and $\theta$
be the angle between them. The general case phase space is then:
\begin{eqnarray}
&&\Phi(l_3, x_3, l_T, \cos\theta, p_T, {M})\equiv\label{PhiGeneral}\\
&&\left(-2 \,{l_T} ({\cos\theta }\,
   {p_T}+{l_T})+2\, {l_3}
   {x_3}+M^2\right)^2\nonumber \\
&& -4
   \left({l_3}^2+{l_T}^2\right)
   \left(2\, {\cos\theta }\, {l_T}\,
   {p_T}+{l_T}^2+{p_T}^2+{x_3}^2\right)=0\nonumber
\end{eqnarray}
for which the generalization of the $p_T=0$ result of Eq.~(\ref{x3solution})
 is 
        \begin{eqnarray}
 &&    x_3^\pm(M,l_3,\cos\theta,p_T)=\\
&&{l_3\over M^2} \left[M^2+2\, {p_T}
   \left({p_T}\pm{\cos\theta }
   \sqrt{M^2+{p_T}^2}\right)
   \right]\nonumber
   \label{fullx3}
   \end{eqnarray}
and that of $|l_T|<M/2$ is
\begin{equation}
{l_T}^{\rm max}(M,\cos\theta,p_T)=\frac{M^2/2}{\sqrt{M^2+p_T^2}+ p_T \cos
   (\theta )}
\end{equation}

The statistically optimal $\Sigma_A$ is computed exactly as in the previous
section, with the result:
\begin{equation}
\Sigma_A(l_3, x_3, l_T, \cos\theta, p_T, {\cal M})={
l_T-{l_T}^{\rm max}({\cal M})\over n_1({\cal M})}
\label{SigmaAgen}
\end{equation}
where $n_1$ is computed as in Eq.~(\ref{n}) in terms of the
phase space function of Eq.~(\ref{PhiGeneral}). More explicitly:
   \begin{eqnarray}
  N_1&=&
   -4\, \big[p_T \cos (\theta ) \left(2 l_3 W_3+{\cal M}^2\right)\nonumber\\
   &+&2 \,l_T \left({\cal M}^2+p_T^2\sin^2(\theta)+W_3^2\right)\big]
   \nonumber\\
   N_2 &=&-4 \left(l_3 {\cal M}^2+2 l_3 p_T^2+2 W_3
   l_T^2-{\cal M}^2 W_3\right)\nonumber\\
   &-& 8 l_T
   \left(l_3+W_3\right) p_T \cos (\theta )\nonumber\\
   N_3&=&4 l_3 \left({\cal M}^2 -2 l_T
   p_T \cos (\theta ) \right)
   - 8  l_T^2 W_3
   \end{eqnarray}
 Some examples of the general phase space surface are given in
Fig.~\ref{fig:PhaseSpaceThetapT}.
 
   \begin{figure}[htbp]
\begin{center}
\includegraphics[width=0.45\textwidth]{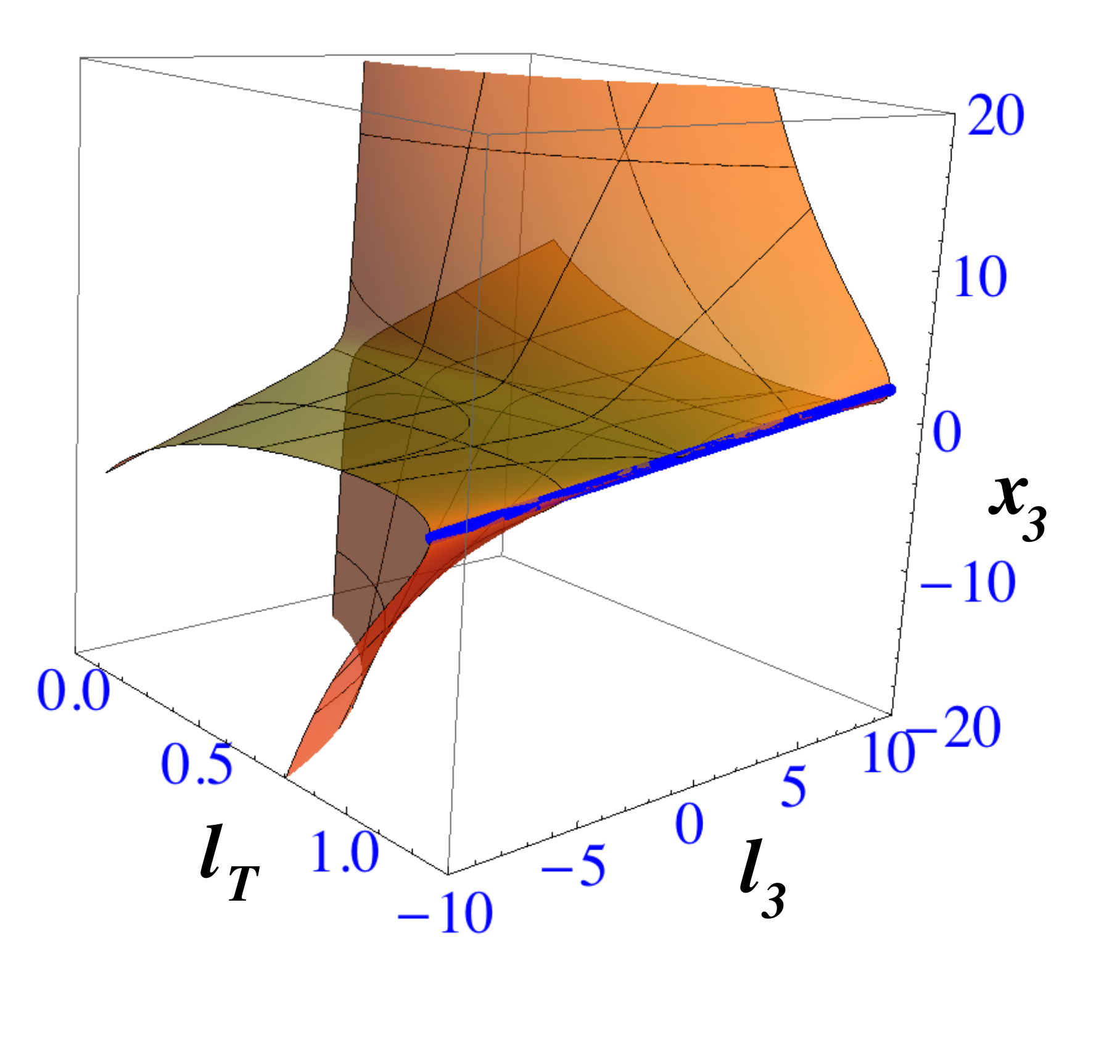}\\
\vspace{.3cm}
\includegraphics[width=0.45\textwidth]{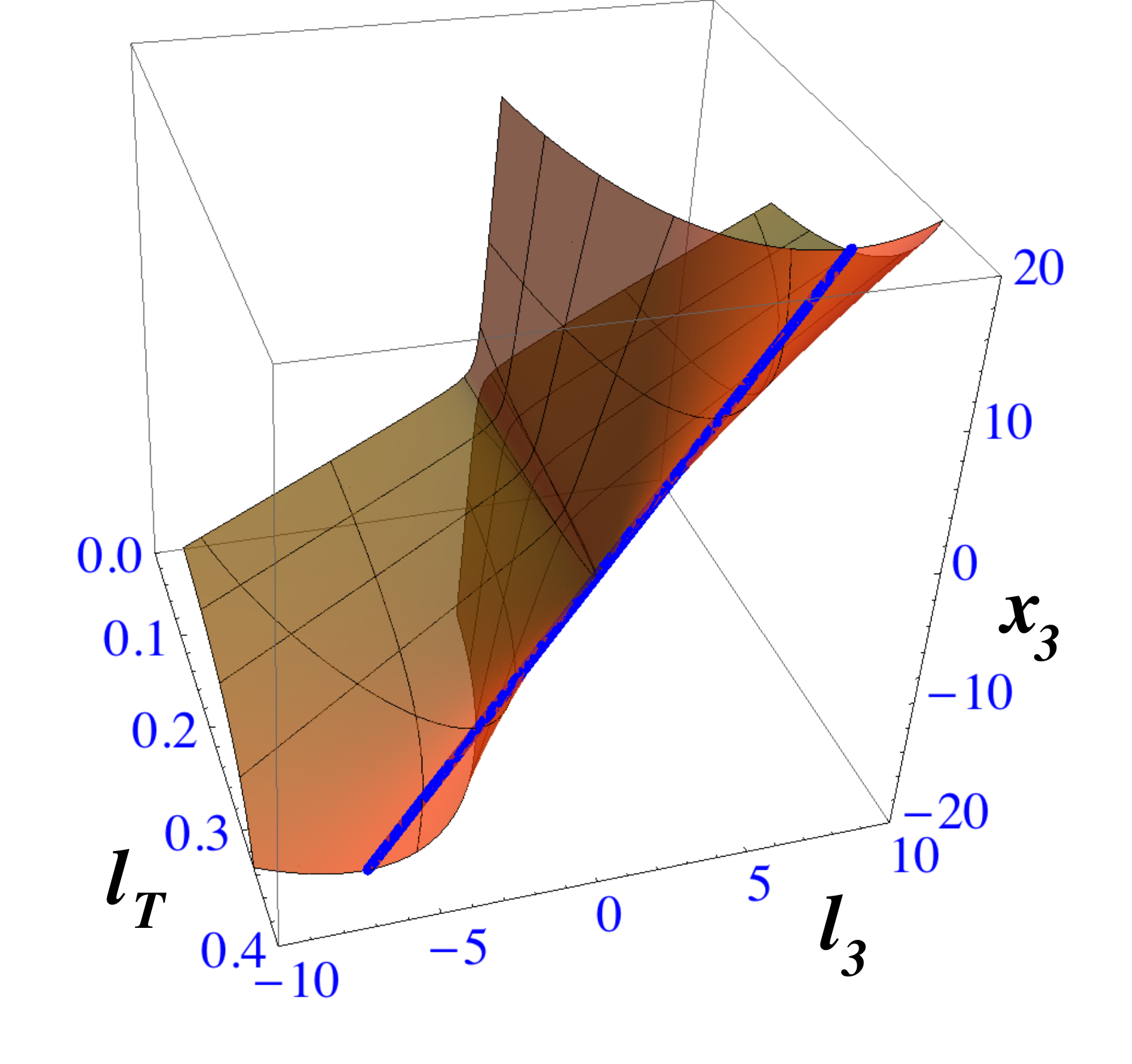}\\
\vspace{.2cm}
\includegraphics[width=0.45\textwidth]{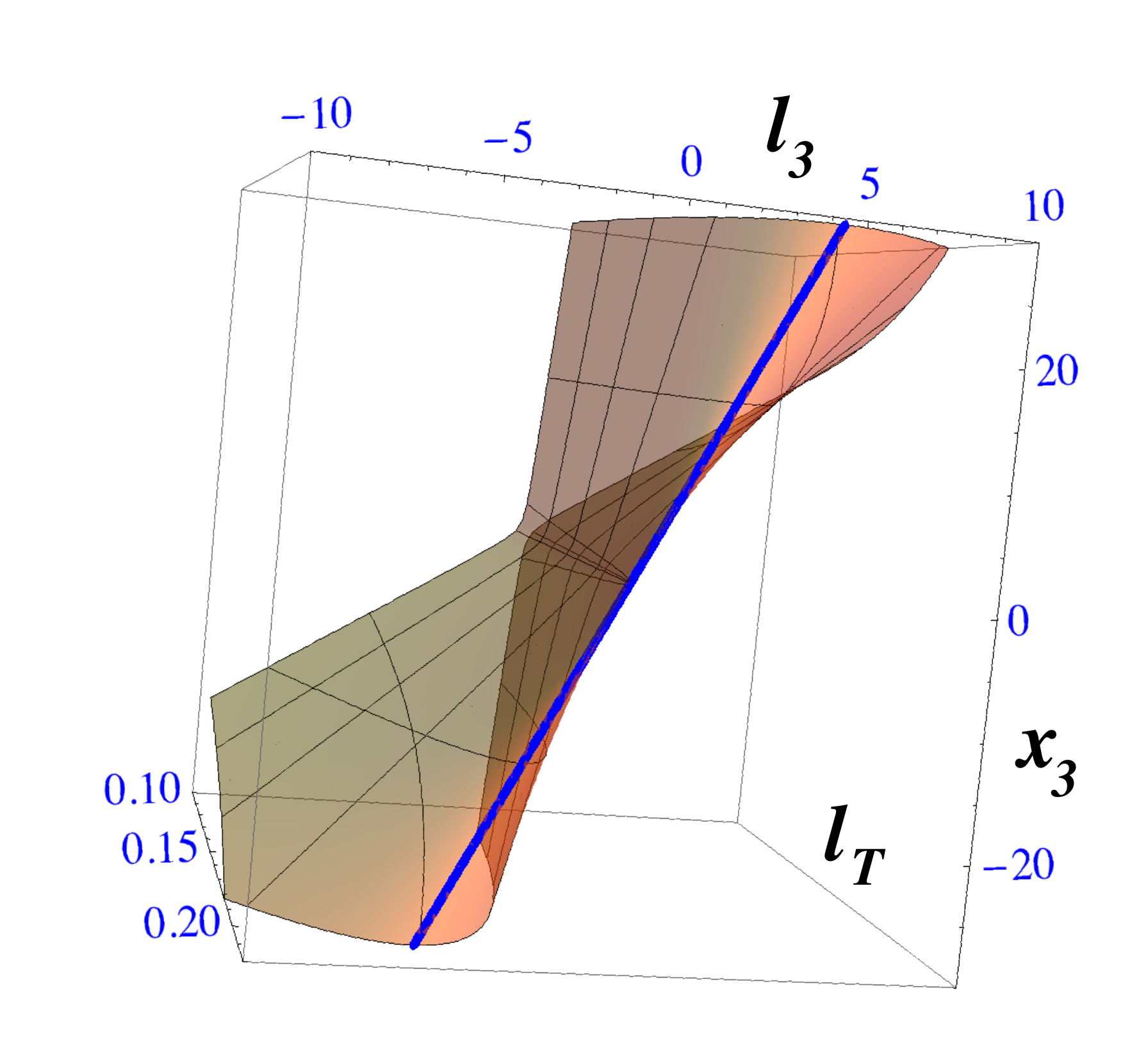}\\
\caption{The general phase space of Eq.~(\ref{PhiGeneral})
for $M=1$ and $p_T=1$. Top, Center, Bottom are
for $\cos\theta=-1,0,1$.
  \label{fig:PhaseSpaceThetapT}}
\end{center}
\end{figure}

\section{Conclusions and outlook}

We have studied in detail the phase space of the simplest
interesting hadron collider process involving an unobservable particle
and only one mass to be determined.
Naturally, the crucial ingredients are the phase space projections
onto the observable momenta, their limits, and the distances of
actual events from these limits.

The edge of the projected phase space is given by the formal singularity 
condition, Eq.~(\ref{Det}), which can be re-expressed as a function of the 
observable momenta, Eq.~(\ref{mT2bis}) and coincides with the
consuetudinary transverse mass function, Eq.~(\ref{mT2}).

The ``singularity variables" are various measures of the distance of an actual
event to the nearest edge singularity. 
We have determined in \S\ref{sec:QOV} the measure for which SV is statistically optimal, 
which we
called the ``statistical squared derivative" and turns out to be well known to statisticians
as the ``Fisher information" \cite{mmm}. The actual result ought to have been
obvious for starters: the optimal variable --$\Sigma_A$ in Eqs.~(\ref{SigmaA},\ref{SigmaAgen})--
is orthogonal to the phase space at all points and is thereby most sensitive to the
unknown mass, which determines the overall scale of momenta.

 Somewhat unexpectedly, singularity variables other than the optimal one develop
 fake singularities away from the edge singularity at $\sigma=0$, see Fig.~(\ref{fig:KimDist}), 
 top. The $W$'s natural width suffices to merge the edge and fake singularities, resulting
 in a peak at $\sigma>0$, see Fig.~(\ref{fig:KimDist}), bottom. This is
 a potential complication in their use as tools to determine the unknown mass(es).

Contrary to the SCs, the SVs depend on longitudinal momenta. In the case of single-$W$
production, whether or not they may add significant precision to a measurement of the $W$ mass
depends on the prior level of understanding of the relevant pdfs \cite{Dydak}, 
a question that we have not
tried to investigate. It may well turn out, contrariwise, that the optimal SV, with a
value of ${\cal M}$ determined by the transverse observables, is a good tool to
constrain the pdfs. 

The SVs contain the SC as a factor. 
This makes them ``weak", in that they are highly
correlated to the information contained in the SC, as discussed in \S\ref{sec:C}.
The SVs are functions of an auxiliary mass
$\cal M$, and of transverse and longitudinal momenta. Varying $\cal M$ as in 
the lower Fig.~(\ref{fig:SingleWSingCond}) is an efficient way to ``focus" on the
relevant mass scale, particularly for cases with more than one unknown mass \cite{Kim}.
But it does not add to the precision with which the mass(es) may be measured.

Whether or not the various and rather negative conclusions of the previous 
two paragraphs apply
to cases wherein more than one particle decays into invisible ones is a question
that we plan to discuss in subsequent work. The answer requires a detailed study
of the relevant phase space, akin to the one in this note.

\vspace{1cm}

{\bf Acknowledgements} 

We are indebted to Frederik Dydak, Francisco
Javier Gir\'on, Ben Gripaios, Cayetano Lopez, Rakhi Mahbubani, 
Maurizio Pierini, Chris Rogan and Raymond Stora for comments and discussions.

\end{document}